\documentclass[preprint,secnumarabic,amssymb, nobibnotes, aps, prd]{revtex4-1}

\usepackage{graphicx,footmisc}
\usepackage{subfig}
\usepackage{natbib,url}
\usepackage{lineno,amsmath,amssymb}
\newcommand*{\be}{\begin{equation}}
\newcommand*{\ee}{\end{equation}}

\def\begineq{\begin{equation}}
\def\endeq{\end{equation}}

\def\begineqn{\begin{equation*}}
\def\endeqn{\end{equation*}}

\def\beginar{\begin{eqnarray}}
\def\endar{\end{eqnarray}}
\def\beginarn{\begin{eqnarray*}}
\def\endarn{\end{eqnarray*}}

\def\lb{\left ( }
\def\rb{\right ) }
\def\lsq{\left [ }
\def\rsq{\right ] }
\def\lbr{\left \langle }
\def\rbr{\right \rangle }

\def\ep{\epsilon}

\def\Rat{\widetilde{Ra}}
\def\Rmt{\widetilde{Rm}}
\def\Pmt{\widetilde{Pm}}
\def\Ltc{\widetilde{L}_c}
\def\kt{\widetilde{k}}
\def\omt{\widetilde{\omega}}

\def\ub{\mathbf{u}}
\def\He{\overline{\mathcal{H}}}
\def\Bb{\mathbf{B}}
\def\ubp{\mathbf{u}^{\prime}}
\def\bbp{\mathbf{b}^{\prime}}

\def\mBb{\overline{\bf B}}
\def\mJb{\overline{\bf J}}

\def\mth{\overline{\vartheta}}
\def\pth{\vartheta^{\prime}}

\def\mBx{\overline{B}_x}
\def\mBy{\overline{B}_y}

\def\pbx{b^{\prime}_x}
\def\pby{b^{\prime}_y}
\def\pbz{b^{\prime}_z}

\def\pjb{\mathbf{j}^{\prime}}
\def\pjx{j^{\prime}_x}
\def\pjy{j^{\prime}_y}
\def\pjz{j^{\prime}_z}

\def\dt{{\partial_{T}}}
\def\dtau{{\partial_{\tau}}}

\def\dsx{{\partial_x}}
\def\dsy{{\partial_y}}

\def\dst{{\partial_t}}

\def\dsz{{\partial_z}}

\def\dz{{\partial_Z}}
\def\dzt{{\partial^2_Z}}

\def\hz{{\bf\widehat z}}

\def\hy{{\bf\widehat y}}
\def\hx{{\bf\widehat x}}

\def\litx{{\bf x}}

\def\emf{\overline{\boldsymbol{\mathcal{E}}}}

\def\md{D^\perp_t}

\def\lp{{\nabla_\perp^2}}
\def\lpi{{\nabla_\perp^{-2}}}

\begin{document}

\title{Convection-driven kinematic dynamos at low Rossby and magnetic Prandtl numbers}%

\author{Michael A. Calkins$^1$\email[corresponding author: ]{michael.calkins@colorado.edu}, 
Louie Long$^2$,
David Nieves$^2$,
Keith Julien$^2$, and
Steven M. Tobias$^3$}%

\affiliation{
$^1$Department of Physics, University of Colorado, Boulder, CO  80309, USA \\
$^2$Department of Applied Mathematics, University of Colorado, Boulder, CO  80309, USA \\
$^3$Department of Applied Mathematics, University of Leeds, Leeds, UK LS2 9JT}

\begin{abstract}
Most large-scale planetary magnetic fields are thought to be driven by low Rossby number convection of a low magnetic Prandtl number fluid. 
Here kinematic dynamo action is investigated with an asymptotic, rapidly rotating dynamo model for the plane layer geometry that is intrinsically low magnetic Prandtl number. The thermal Prandtl number and Rayleigh number are varied to illustrate fundamental changes in flow regime, ranging from laminar cellular convection to geostrophic turbulence in which an inverse energy cascade is present. A decrease in the efficiency of the convection to generate a dynamo, as determined by an increase in the critical magnetic Reynolds number, is observed as the buoyancy forcing is increased. This decreased efficiency may result from both the loss of correlations associated with the increasingly disordered states of flow that are generated, and boundary layer behavior that enhances magnetic diffusion locally. We find that the spatial characteristics of $\alpha$, and thus the large-scale magnetic field, is dependent only weakly on changes in flow behavior. However, our results are limited to the linear, kinematic dynamo regime, and future simulations including the Lorentz force are therefore necessary to assess the robustness of this result. In contrast to the large-scale magnetic field, the behavior of the small-scale magnetic field is directly dependent on, and therefore shows significant variations with, the small-scale convective flow field. 

\end{abstract}

\maketitle

\section{Introduction}

Most planets within the Solar System possess large-scale magnetic fields that are thought to be the result of dynamo action occurring within their electrically conducting fluid interiors \citep{gS11}. Detailed spatiotemporal observations of these systems are lacking, therefore placing emphasis on the development of theory and models for understanding the physical mechanisms responsible for the generation of planetary magnetic fields \citep[e.g.][]{sT11}. Generally speaking, modeling geo- and astrophysical fluids is difficult due to the broad range of spatial and temporal scales that must be resolved. Although significant advances in the understanding of planetary dynamos have been made by direct numerical simulation (DNS) of the full set of governing magnetohydrodynamic equations, accessing planetary-like flow regimes with DNS is currently impossible with modern-day high-performance computing environments \citep{cJ11b}. An alternative, but complementary strategy to DNS, is the development of reduced models that filter dynamically unimportant phenomena and result in simplified governing equations. 
The benefit of reduced models is that they are significantly more efficient to solve numerically given that various terms from the equations are systematically removed or simplified. As a result of these simplifications, reduced models can provide enhanced physical interpretation and insight into observed hydrodynamic and magnetohydrodynamic processes. In the present work we utilize an asymptotic model recently developed by \citeauthor{mC15b} to examine the behavior of convection-driven kinematic dynamo action in the limit of rapid rotation.

It is widely believed that most planetary dynamos are powered by buoyancy-driven motions associated with the convective transport of heat and chemicals. Moreover, it is thought that the Coriolis force constrains fluid motions within the interiors of most planets given that the observed large-scale magnetic fields are predominantly aligned with the rotation axis of the planet \citep{gS11}. Indeed, it was recognized early on in the seminal studies of \citet{eP55} and \citet{mS66a} that the Coriolis force provides a robust mechanism for breaking the reflectional up/down symmetry of fluid motions that is necessary for large-scale magnetic field generation. 

The rotating convection dynamics of an Oberbeck-Boussinesq fluid depend upon three non-dimensional parameters
\be
Ra = \frac{g \gamma \Delta T H^3}{\nu \kappa} , \quad Ek = \frac{\nu}{2 \Omega H^2}, \quad Pr = \frac{\nu}{\kappa},
\ee
where the Rayleigh number $Ra$ controls the strength of the buoyancy force, the Ekman number $Ek$ is the ratio of  viscous and Coriolis forces, and the thermal Prandtl number $Pr$ is the ratio of viscous to thermal diffusivity.  Here $H$ is the depth of the fluid layer and $\Delta T$ represents the (constant) temperature difference between the lower and upper solid boundaries.  The fluid is characterized by thermal expansion coefficient $\gamma$, kinematic viscosity $\nu$ and thermal diffusivity $\kappa$.  The rotation rate of the system is assumed constant and parallel to the axial direction $\hz$, i.e.~$\mathbf{\Omega} = \Omega \hz$.

The rapidly rotating convection regime is characterized by the asymptotic limits $Ek \rightarrow 0$ and $Pr \gg Ek$, with $Ra = O(Ek^{-4/3})$ \citep{sC61}. In this regime, the Coriolis and pressure gradient forces are in approximate balance, resulting in quasi-geostrophic (QG) dynamics. Linear theory has established that the horizontal wavelength in QG convection scales as $L \sim H Ek^{1/3}$, such that the aspect ratio $H/L = Ek^{-1/3} \gg 1$. Provided the Rossby number 
\be
Ro = \frac{U}{2 \Omega L},
\ee 
is also small, these scalings also hold in the nonlinear regime \citep[e.g.][]{kJ98a}. Here $U$ is a characteristic velocity scale of the convection and the Rossby number characterizes the ratio of inertial forces to the Coriolis force. If we employ a viscous scaling for the velocity $U=\nu/L$, we have
\be
\ep \equiv Ro = Ek^{1/3},
\ee 
where we have utilized the aspect ratio scaling previously mentioned.

\citet{sC72} exploited the asymptotic properties of QG convection to develop the first reduced dynamo model for rotating convection. They focused on small amplitude weakly nonlinear motions in which the small-scale (convective) Reynolds number
\be
Re = \frac{U L}{\nu},
\ee
is small, and the magnetic Prandtl number
\be
Pm = \frac{\nu}{\eta},
\ee
where $\eta$ is the magnetic diffusivity, is order unity. Dynamo action was investigated in detail by \citet{aS74} and \citet{yF82} for magnetic fields of varying strengths within the context of the Childress-Soward dynamo model. It was shown that QG convection acts as an efficient magnetic field generator in the sense that growing magnetic fields are observed just above the onset of convection with a large-scale magnetic field that oscillates in time. 


The quasi-geostrophic dynamo model (QGDM) recently developed by \citet{mC15b} is a fully nonlinear generalization of the Childress-Soward dynamo model, and is valid in the limit $(Ro,Ek)\rightarrow 0$ and $Re = O(1)$.  The QGDM can thus be used to explore both linear (i.e.~kinematic) and nonlinear dynamo behavior for all rapidly rotating convection regimes from near the onset of convection to strongly forced, convective turbulence. Distinct versions of the QGDM were developed for both $Pm=O(1)$ and $Pm \ll 1$, and it was shown that the Lorentz force and the reduced induction equations have unique forms for these two cases.  \citet{mC16} used the QGDM to investigate the kinematic problem, in which the Lorentz force in the momentum equations is neglected, utilizing the single wavenumber solutions considered by \citet{aB94} and \citet{kJ98,kJ99b}.  It was shown that low $Pm$ dynamos are readily attainable for low $Ro$ convection, with low $Pr$ convection the most efficient given the reduced critical Rayleigh number in comparison to high $Pr$ fluids.  

We stress here that the characteristics of kinematic dynamo action depends crucially on the value of $Pm$ \citep[][]{sT11}; for rapidly-rotating flows there is also a sensitivity to the Ekman number \citep[e.g.][]{nS06}. \citet{eD16} argues that a distinguished limit should be sought where both $Pm$ and $Ek$ are small with $1 \gg Pm \gg Ek$ for the dynamics to begin to replicate that of planetary interiors. As shown by \citet{mC15b}, the dynamo utilized here, in contrast with virtually all numerical models of planetary dynamos, does satisfy this inequality.

In the absence of magnetic field, the QGDM is equivalent to the QG convection equations first developed by \citet{kJ98a} and extensively explored via simulations in \citet{mS06} and \citet{kJ12}. Neglecting the buoyancy force and dissipation, these equations are also mathematically identical to those developed by \citet{kS79} in their investigation of low frequency (i.e.~geostrophic) inertial waves in deep fluid layers \citep[see also][]{sN11}.  The linear, spherical convection investigations of \citet{pR68} and \citet{fB70}, and later work by \citet{cJ00} and \citet{eD04}, utilized a mathematically identical approach that exploited a leading order geostrophic balance and spatial anisotropy.  Moreover, reduced QG convection equations were recently developed for the three-dimensional cylindrical annulus geometry \citep{mC13}, extending the small-slope two dimensional model first developed by \citet{fB70}. Collectively, these previous investigations highlight the importance of developing and employing asymptotic models for the purpose of improving our understanding of flow regimes that are characteristic of planets.

Simulations of the QG convection equations have identified four primary flow regimes in rapidly rotating, non-magnetic Rayleigh-B\'enard convection \citep{mS06,kJ12}. These regimes can be identified by the predominance of a given flow morphology, and can be referred to as the (1) cellular, (2) convective Taylor column, (3) plume and (4) geostrophic turbulence regimes.  Each of these regimes is characterized by unique heat transfer behavior and flow statistics \citep{kJ12b,dN14}. The final geostrophic turbulence regime is dominated by an inverse cascade that generates a depth invariant dipolar vortex which fills the horizontal extent of the computational domain \citep{aR14}. An identical inverse cascade mechanism was identified previously in stochastically-forced, rotating homogeneous turbulence \citep{lS96,lS99}, and subsequent DNS studies have demonstrated it as a robust phenomenon in rapidly rotating convection \citep{sS14,bF14,cG14}. DNS investigations \citep{sS14} and laboratory experiments \citep{jmA15,jC15} show excellent agreement with the simulations of the QG convection equations. In particular, \citet{sS14} showed that it is necessary to reach very small Ekman numbers ($Ek \lesssim 10^{-7}$) and Rossby numbers ($Ro \lesssim 0.05$) to reach the asymptotic regime in which the dipolar structure of the inverse-cascade-generated vortex is preferred over the cyclonic vortices that become predominant in the broken-symmetry regime present at higher Ekman and Rossby numbers \citep{lS99,pV02,bF14,cG14}.

Many previous DNS dynamo investigations have been undertaken in the Rayleigh-B\'enard, plane layer geometry \citep{cJ00b,jR02,sS04,fC06,sT08,aT12,aT14,cG15}.  Investigations in spherical geometries are of obvious importance for planets, but must employ lower efficiency numerics typically and are therefore more restricted in parameter space \citep{jmA15}.  The DNS investigation of \citet{sS04} confirmed the predictions of \citet{aS74} that a strong, time-oscillatory mean magnetic field can be generated near the onset of convection.  \citet{aT12,aT14} found a transition from the mean field dynamo mechanism of \citet{sC72} to a dynamo of fluctuation-type.  The recent work of \citet{cG15} examined the influence of the inverse cascade and associated large-scale vortex on the resulting dynamo; they found that for $Pm \gtrsim 1$, the mean magnetic field is weak and the inverse cascade is less pronounced, in comparison to a case at $Pm = 0.2$ where a significant inverse cascade and mean magnetic field are present. It should be noted that all of these investigations, despite the advantage of being in a computationally simple plane layer, are at parameters well away from those relevant to planetary interiors.

In the present work we extend the single mode kinematic investigation of \citet{mC16} to the fully nonlinear, multi-mode case by performing numerical simulations of the QG convection equations in which all of the dynamically active scales are present. As mentioned previously, these equations have been studied in some detail; our main focus is utilizing the simulations to collect the necessary statistics for studying the onset of dynamo action.  Moreover, we utilize two different Prandtl numbers to illustrate the differences between dynamos driven by thermal and compositional convection.

\section{The quasi-geostrophic dynamo model (QGDM)}
\label{S:eqn}

A brief overview of the derivation and main features of the QGDM are given in the present section; for a detailed discussion the reader is referred to \citet{mC15b}.  The basic premise of the QGDM is similar to the Childress-Soward dynamo model in that we exploit the anisotropic structure of the convection, where $H/L = \ep^{-1}$. This characteristic anisotropy in rotating convection motivates the use of multiple scale asymptotics in the space (along the axial direction) and time dimensions, and results in the expansions \citep[e.g.][]{cB10}
\be
\dsz \rightarrow \dsz + \ep \dz , \quad
\dst \rightarrow \dst + \ep^{3/2} \dtau + \ep^2 \dt ,
\ee
where $Z = \ep z$ is the large-scale vertical coordinate over which convection occurs, $\tau = \ep^{3/2} t$ is the slow mean magnetic field timescale, while $T=\ep^2 t$ is the slow mean temperature timescale.  The slow and fast independent variables are therefore denoted by $(Z,\tau,T)$ and $\lb \litx,t \rb$, respectively.  Slow horizontal scales can also be utilized, though we neglect this effect in the present work \citep[for a more general discussion see][]{mC15b}.  

All of the dependent variables are decomposed into mean and fluctuating variables according to
\be
f(\litx,Z,\tau,T) = \overline{f}(Z,\tau,T) + f'(\litx,Z,\tau,T),
\ee
with the fast averaging operator defined by
\be
\overline{f}(Z,\tau,T)  = \lim_{t', \mathcal{V} \rightarrow \infty} \, \frac{1}{t' \mathcal{V}} \int_{t',\mathcal{V}} f(\litx,Z,t,\tau,T) \, d \mathbf{x} \, dt , \quad \overline{f'} \equiv 0 ,
\ee
where $\mathcal{V}$ is the small-scale fluid volume. Each variable is then expanded as a power series according to
\be
\begin{split}
f(\litx,Z,t,\tau,T)  = \overline{f}_0(Z,\tau,T) + f'_0(\litx,Z,t,\tau,T) + \\\ep^{1/2} \lsq \overline{f}_{1/2}(Z,\tau,T) + f'_{1/2}(\litx,Z,t,\tau,T) \rsq + \\ \ep \lsq \overline{f}_{1}(Z,\tau,T) + f'_{1}(\litx,Z,t,\tau,T) \rsq + O(\ep^{3/2}) . \label{E:expand}
\end{split}
\ee
We substitute the above expansions for each variable into the governing equations, separate into mean and fluctuating equations, and collect terms of equal asymptotic order. By construction, the leading-order balance in the fluctuating momentum equation is geostrophy
\be
\hz \times \ubp_{0,\perp} = - \nabla_\perp p'_1 ,
\ee
where the (fluctuating) geostrophic velocity is $\ubp_{0,\perp} = (u'_0,v'_0,0)$, $p'_1$ is the fluctuating pressure and $\nabla_\perp = (\dsx,\dsy,0)$.  Mass conservation at leading-order is horizontally non-divergent
\be
 \nabla_\perp \cdot \ubp_0=0,
\ee
which allows us to define the geostrophic streamfunction $\ubp_{0,\perp} = - \nabla \times \psi'_0 \hz$ and the corresponding axial vorticity $\zeta'_0 = \lp \psi'_0$.  Vortex stretching is captured via mass conservation at order $\ep$, 
\be
\nabla\cdot \ubp_1 + \dz w'_0 = 0,
\ee
where the higher order ageostrophic velocity field is given by $\ub'_1$.

The prognostic momentum equation is obtained at $O(1)$ in the asymptotic expansion. Closure of this equation is obtained by imposing solvability conditions; the end result is that the three components of the momentum equation are reduced to a vertical vorticity equation and a vertical momentum equation that do not depend upon the small-scale vertical coordinate $z$. Similar reductions are performed for the heat and magnetic induction equations.

Hereafter, we drop subscripts denoting asymptotic ordering.  The complete set of reduced equations, non-dimensionalized using the small-scale viscous diffusion time $L^2/\nu$, is given by 
\begin{gather}
\md \lp \psi' - \dz w' = \mBb \cdot \nabla_\perp \pjz + \nabla_\perp^4 \psi', \label{E:vort0} \\
\md w' +  \dz \psi' = \frac{\Rat}{Pr}  \pth  + \mBb \cdot \nabla_\perp \pbz + \lp w', \label{E:mom0} \\
 \md \pth +  w \dz \mth = \frac{1}{Pr} \lp \pth , \\
 \dt \mth + \dz \overline{\lb w' \pth \rb} =  \frac{1}{Pr} \dzt \mth , \label{E:mheat} \\
  \partial_\tau \mBb =  \hz \times \dz \emf  + \frac{1}{\Pmt} \dzt \mBb, \label{E:minduc1}\\
  0 = \mBb \cdot \nabla_\perp \ubp + \frac{1}{\Pmt} \lp  \bbp \label{E:finduc1} .
\end{gather}
Here the temperature and magnetic field vector are given by $\vartheta = \mth + \ep \pth$ and $\Bb = \mBb + \ep^{1/2} \bbp$, respectively.  The components of the mean and fluctuating magnetic field vectors are denoted by $\mBb = (\mBx,\mBy,0)$ and $\bbp = (\pbx,\pby,\pbz)$ and the corresponding fluctuating current density is $\pjb = (\pjx,\pjy,\pjz) = (\dsy \pbz,-\dsx \pbz, \dsx \pby - \dsy \pbx)$.  The mean electromotive force (emf) is denoted by $\emf = \overline{\lb \ubp \times \bbp \rb}$.

The reduced Rayleigh number and the reduced magnetic Prandtl number appearing in the above system of equations are defined by
\be
\Rat = \ep^4 Ra, \quad Pm = \ep^{1/2} \Pmt .
\ee
The critical reduced Rayleigh number, wavenumber and frequency characterizing the onset of convection are denoted as $\Rat_c$, $\kt_c$ and $\omt_c$. The onset of convection is steady ($\omt_c=0$) for $Pr \gtrsim 0.68$, in which case $\Rat_c \approx 8.6956$ and $\kt_c \approx 1.3048$ \citep{sC61,kJ98}. In the present work we discuss results for $Pr=1$ and $Pr=10$. 

The boundary conditions for the reduced system are impenetrable, stress-free, fixed-temperature and electrically conducting,
\be
w' = 0 \quad \textnormal{at} \quad Z = 0, 1,\label{E:vbc}
\ee
\be
\mth = 1\quad \textnormal{at} \quad Z = 0, \quad \textnormal{and} \quad 
\mth = 0\quad \textnormal{at} \quad Z = 1 , \label{E:tbc}
\ee
\be
\dz \mBb = 0,\quad \textnormal{at} \quad Z = 0, 1. \label{E:mbc}
\ee
We note that neither the thermal boundary conditions nor the magnetic boundary conditions influence the main results of the present work. In the limit of rapid rotation, fixed temperature and fixed heat flux thermal boundary conditions become equivalent provided that no large-scale horizontal modulation is present  \citep{mC15c}. Moreover, for the present kinematic dynamo problem both perfectly conducting and perfectly insulating electric boundary conditions yield identical stability criteria \citep[e.g.~see][]{bF13c}.

In the present investigation we consider only the kinematic dynamo problem in the sense that the Lorentz terms appearing in equations \eqref{E:vort0}-\eqref{E:mom0} are ignored. This is done to determine the region of parameter space for which exponentially growing magnetic fields are present. The equations are therefore linear in the magnetic field vectors $\mBb$ and $\bbp$, whose solutions can be sought via an eigenvalue formulation of equations \eqref{E:minduc1}-\eqref{E:finduc1} and \eqref{E:mbc} once $\ubp$ is determined from equations \eqref{E:vort0}-\eqref{E:mheat} and \eqref{E:vbc}-\eqref{E:tbc}.

With the use of \eqref{E:finduc1} we can eliminate $\bbp$ and the two components of the emf become
\be
\emf_i = \alpha_{ij} \overline{B}_j,
\ee
where the pseudo-tensor $\alpha_{ij}$ is given by
\be
\alpha_{ij} = \Pmt \lb 
\begin{array}{cc} 
\overline{w' \lpi \dsx v' - v' \lpi \dsx w'}  \, & \,  \overline{w' \lpi \dsy v' - v' \lpi \dsy w'} \\
\overline{-w' \lpi \dsx u' + u' \lpi \dsx w'}  &  \overline{- w' \lpi \dsy u' + u'  \lpi \dsy w'}
\end{array}
\rb . \label{E:alphav}
\ee
and $\lpi$ denotes the inverse horizontal Laplacian operator. 

For sufficiently long simulation times, the numerical simulations show that, as expected,  the alpha tensor is diagonal, symmetric and isotropic such that $\alpha_{12} = \alpha_{21} = 0$ and $\alpha_{11} = \alpha_{22} = \alpha$. The induction equations then simplify to become
\be
\partial_\tau \mBx = - \dz \lb \alpha \mBy \rb  + \frac{1}{\Pmt} \dzt \mBx , \label{E:minduc1a}
\ee
\be
\partial_\tau \mBy = \dz \lb \alpha \mBx \rb + \frac{1}{\Pmt} \dzt \mBy , \label{E:minduc1b}
\ee
where the alpha tensor will now be referred to solely by the single pseudo-scalar quanity $\alpha$ hereafter. For specificity, in what follows we refer to $\alpha = \alpha_{11}$ as defined in equation \eqref{E:alphav}, which can also be written more simply as
\be
\alpha = \Pmt \, \overline{ \lb w' \lpi \dsx v' + \dsx v' \lpi w' \rb} . \label{E:alpha}
\ee

The vertically averaged mean magnetic energy equation is given by $\langle \mBb \cdot $\eqref{E:minduc1}$\rangle$ to give
\be
\dtau \lbr \frac{1}{2} \mBb^2 \rbr = \lbr \alpha \mBb \cdot \mJb \rbr - \frac{1}{\Pmt} \lbr \mJb^2 \rbr, \label{E:magen}
\ee
where the angled brackets denote a vertical average and the mean current density is $\mJb = \lb -\dz \mBy , \dz \mBx, 0 \rb$. Marginal stability thus corresponds to $\lbr \alpha \mBb \cdot \mJb \rbr = \Pmt^{-1} \lbr \mJb^2 \rbr$, showing that there must be a non-zero vertically averaged alignment between the two currents $\alpha \mBb$ and $\mJb$ for large-scale  dynamo action to occur, as noted by \citet{hM70b}.

A quantity that is thought to be important in dynamo theory is the (mean) kinetic helicity $\He = \overline{\ubp \cdot \boldsymbol{\zeta}'}$ \citep[e.g.][]{hM78b}, where the asymptotically reduced vorticity vector is given by 
\be
\boldsymbol{\zeta}' = \dsy w' \, \hx - \dsx w' \, \hy + \zeta' \hz.  
\ee
We then have
\be
\He = 2 \, \overline{w' \zeta'} . \label{E:hel}
\ee
For single wavenumber solutions it was shown that $\alpha \propto \He$ \citep{mC16}. However, comparison of equations \eqref{E:alpha} and \eqref{E:hel} shows that $\alpha$ and $\He$ are not as simply related for general, multi-mode convection. Indeed, we show below that the kinetic helicity and $\alpha$ exhibit significant differences in spatial structure and temporal variations, though the magnitudes of the two quantities do show similar trends. 

We also utilize the relative kinetic helicity defined as 
\be
\He_r = \frac{\overline{\ubp \cdot \boldsymbol{\zeta}'}}{\sqrt{\overline{\ubp \cdot \ubp}} \sqrt{\overline{\boldsymbol{\zeta}' \cdot \boldsymbol{\zeta}'}}} .
\ee
A maximally helical flow is one in which $|\He_r| = 1$ in a given region of space.

\subsection{Numerical Methods}

The velocity field is computed by solving equations \eqref{E:vort0}-\eqref{E:mheat} in the absence of the Lorentz terms.  The equations are discretized in the horizontal and vertical dimensions with Fourier series and Chebyshev polynomials, respectively.  The time-stepping is performed with a third order Runge-Kutta scheme developed by \citet{pS91}.  The reader is referred to \citet{mS06} for further details of the numerical methods employed in the simulations. A summary of the numerical simulations used in the present work is given in Table \ref{T:sims}.

\begin{table}
  \begin{center}
    \begin{tabular}{lccccc}
      $Pr$      &     $\Rat$    &     Box Dimensions & $N_x \times N_y \times N_Z$  & $Re$   \\
      \hline      
      \hline
      1           &     10       &   $10\Ltc\times10\Ltc$  &   $64\times64\times33$   & $0.68$      \\      
      1           &     20       &   $10\Ltc\times10\Ltc$  &  $96\times96\times65$    & $3.40$            \\  
      1           &     30       &   $10\Ltc\times10\Ltc$  &  $96\times96\times129$    & $7.01$    \\
      1           &     40       &   $10\Ltc\times10\Ltc$  &  $128\times128\times129$   & $10.4$     \\      
      1           &     50       &   $20\Ltc \times 20\Ltc$  &  $256\times256\times129$  & $13.8$            \\
      1           &     60       &   $20\Ltc \times 20\Ltc$  &  $256\times256\times193$   & $16.6$           \\
      1           &     80       &   $20\Ltc \times 20\Ltc$  &  $384\times384\times193$   & $22.8$            \\
      1           &     100      &   $20\Ltc \times 20\Ltc$  &   $384\times384\times257$  & $31.8$           \\
      10           &     10       &   $10\Ltc\times10\Ltc$  &  $64\times64\times65$    & $0.065$           \\      
      10           &     20       &   $10\Ltc\times10\Ltc$  &   $64\times64\times65$    & $0.29$           \\  
      10           &     30       &   $10\Ltc\times10\Ltc$  &   $96\times96\times97$    & $0.56$           \\
      10           &     40       &   $10\Ltc\times10\Ltc$  &   $96\times96\times97$     & $0.88$           \\      
      10           &     50       &   $10\Ltc\times10\Ltc$  &   $96\times96\times97$    & $1.25$               \\
      10           &     60       &   $10\Ltc\times10\Ltc$  &   $96\times96\times97$    & $1.68$             \\
      10           &     80       &   $10\Ltc\times10\Ltc$  &   $128\times128\times129$    & $2.51$            \\            
      10           &     100      &   $10\Ltc\times10\Ltc$  &   $128\times128\times129$   & $3.34$              \\ 
      10           &     150      &   $10\Ltc\times10\Ltc$  &   $256\times256\times257$   & $5.33$         \\
      10           &     200      &   $10\Ltc\times10\Ltc$  &   $384\times384\times385$    & $6.96$             \\                  
    \end{tabular}
    \caption{Details of the numerical simulations used in the present study.  Here $Pr$ is the Prandtl number, $\Rat$ is the asymptotically scaled Rayleigh number, $N_x = N_y$ is the horizontal spatial resolution, $N_Z$ is the vertical spatial resolution and $Re$ is the small-scale Reynolds number. In each case the non-dimensional horizontal box dimensions are given in integer multiples of the critical horizontal wavelength $\Ltc = 4.1854$.}
    \label{T:sims}
  \end{center}
\end{table}

The horizontal dimensions of the computational domain are specified in terms of multiples of the critical horizontal wavelength of the convection, $\Ltc$; for steady convection $\Ltc = 2 \pi/\kt_c \approx 4.8154$. With respect to collecting useful statistics in as short a wall-clock time as possible (given a statistically stationary convective state), there is a trade-off between the required simulation time and the horizontal dimensions.  For $Pr=10$ cases we found that $10 \Ltc \times 10\Ltc$ dimensions is sufficient to collect converged statistics. For $Pr=1$ and $\Rat < 50$, $10 \Ltc \times 10\Ltc$ dimensions is also sufficient. For $\Rat \ge 50$ the presence of the inverse cascade made simulation times impractical so that the dimensions were increased to $20 \Ltc \times 20 \Ltc$, which greatly accelerated the convergence rate of the statistics.

Solving the generalized eigenvalue problem requires that $\alpha$ is averaged over horizontal planes  for sufficiently long times. A strategy that we employ to speed up the convergence of $\alpha$ to a well-defined statistically steady state is to reflect the profile about the $Z$-midplane ($Z=0.5$), multiply by negative one and take the average of the resulting profiles. In particular, this procedure allows us to obtain an $\alpha$ profile that is perfectly antisymmetric with respect to $Z=0.5$, and therefore more representative of the stationary state.

The generalized eigenvalue problem for the complex eigenvalue $\sigma$ and the mean magnetic field $\mBb$ is solved utilizing MATLAB's sparse eigenvalue solver `sptarn'.  Chebyshev polynomials were used to discretize the vertical derivatives appearing in the governing equations.  We use the same number of Chebyshev polynomials to solve the eigenvalue problem as are used for the numerical simulations (see Table \ref{T:sims}).  To generate a numerically sparse system, we use the Chebyshev three-term recurrence relation and solve directly for the spectral coefficients with the boundary conditions enforced via `tau'-lines \citep{dG93}.  The non-constant coefficient terms appearing in the mean induction equations are treated efficiently by employing standard convolution operations for the Chebyshev polynomials \citep{gB97,sO13}.  An identical approach was used in \citet{mC16} and also for the linear stability of compressible convection \citep{mC15,mC15d}.

\section{Results}

\subsection{Convection characteristics}

\begin{figure}
  \begin{center}
   \subfloat[]{
      \includegraphics[height=4cm]{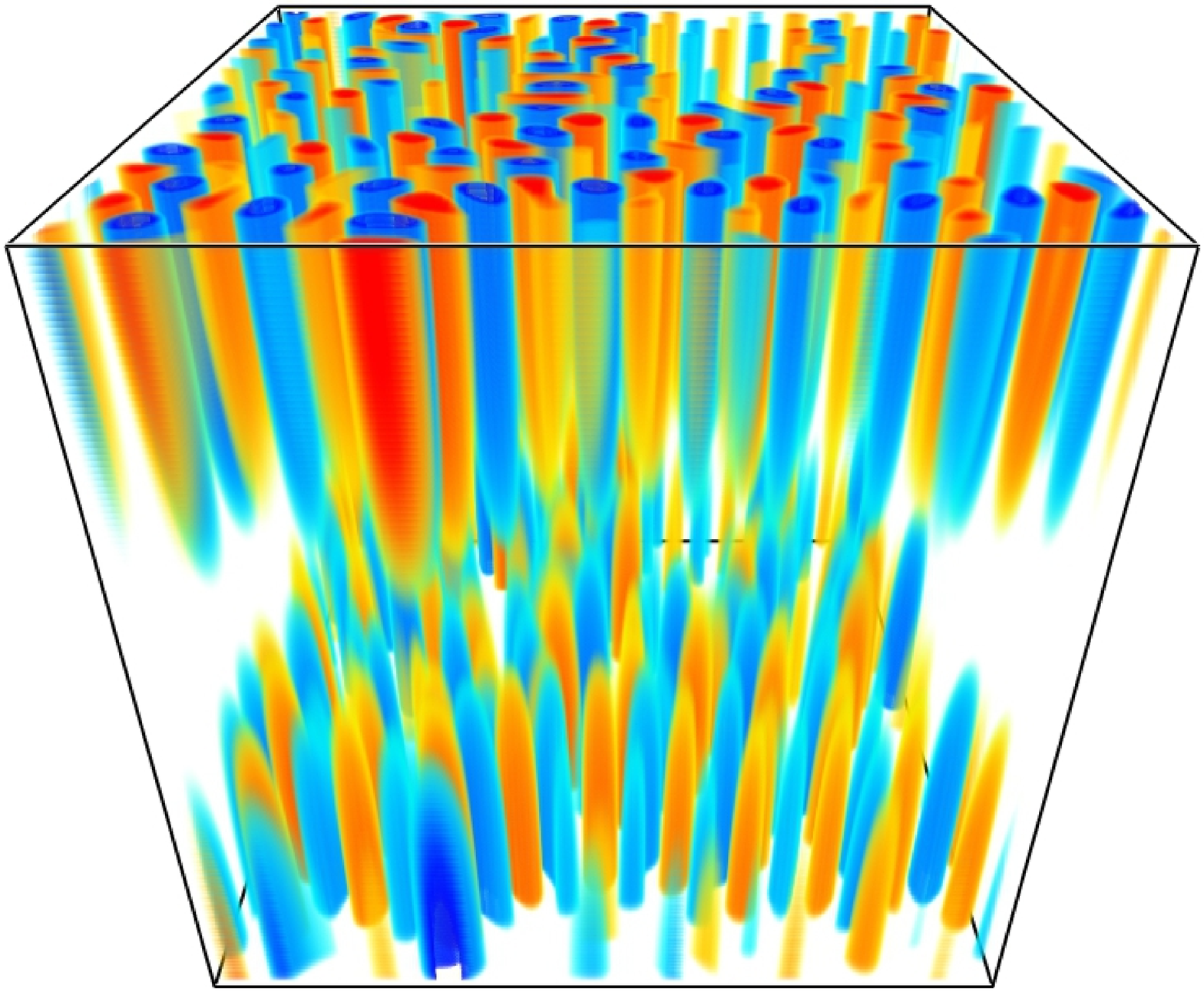}}
      \quad
    \subfloat[]{
      \includegraphics[height=4cm]{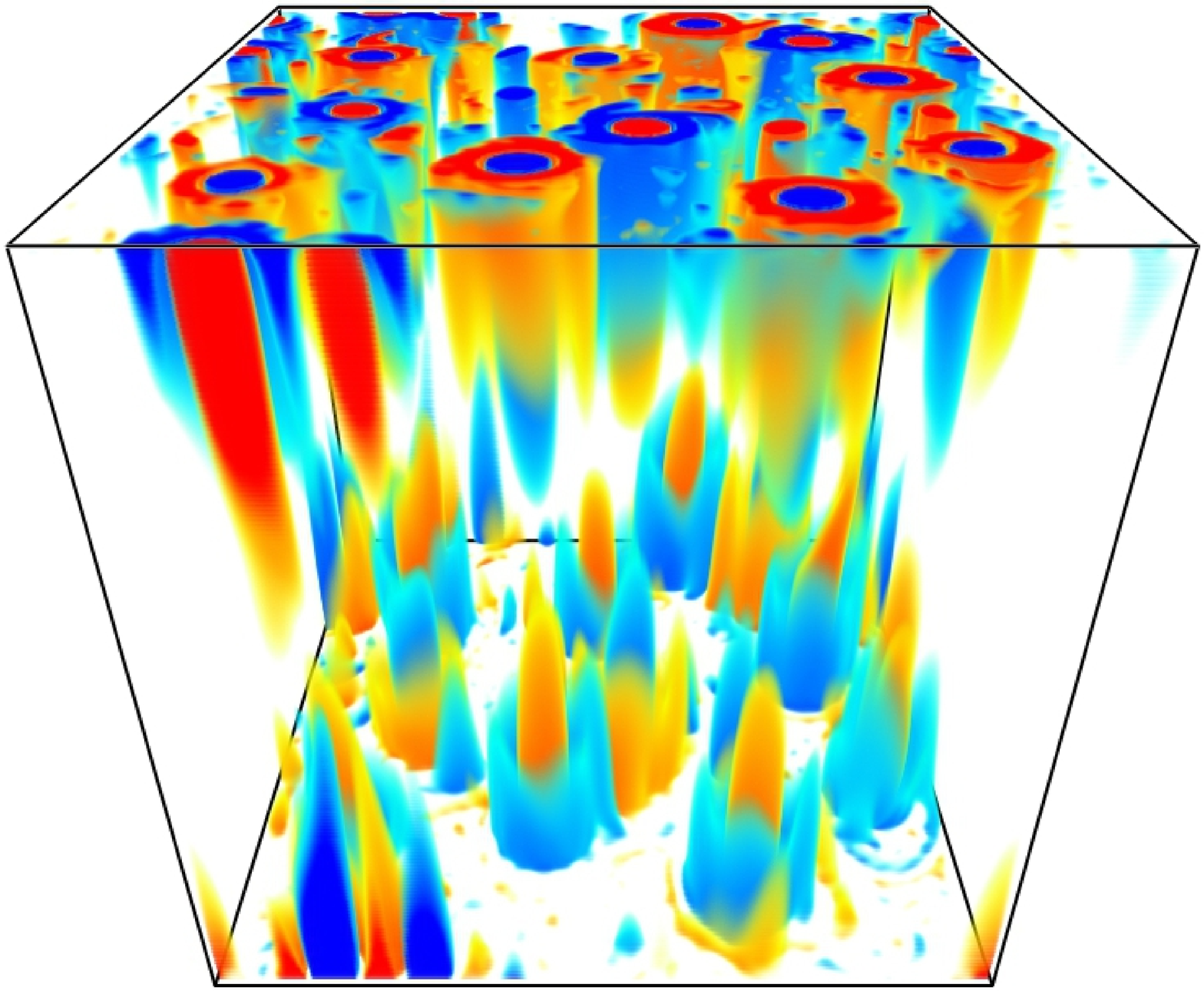}}    \\
   \subfloat[]{
      \includegraphics[height=4cm]{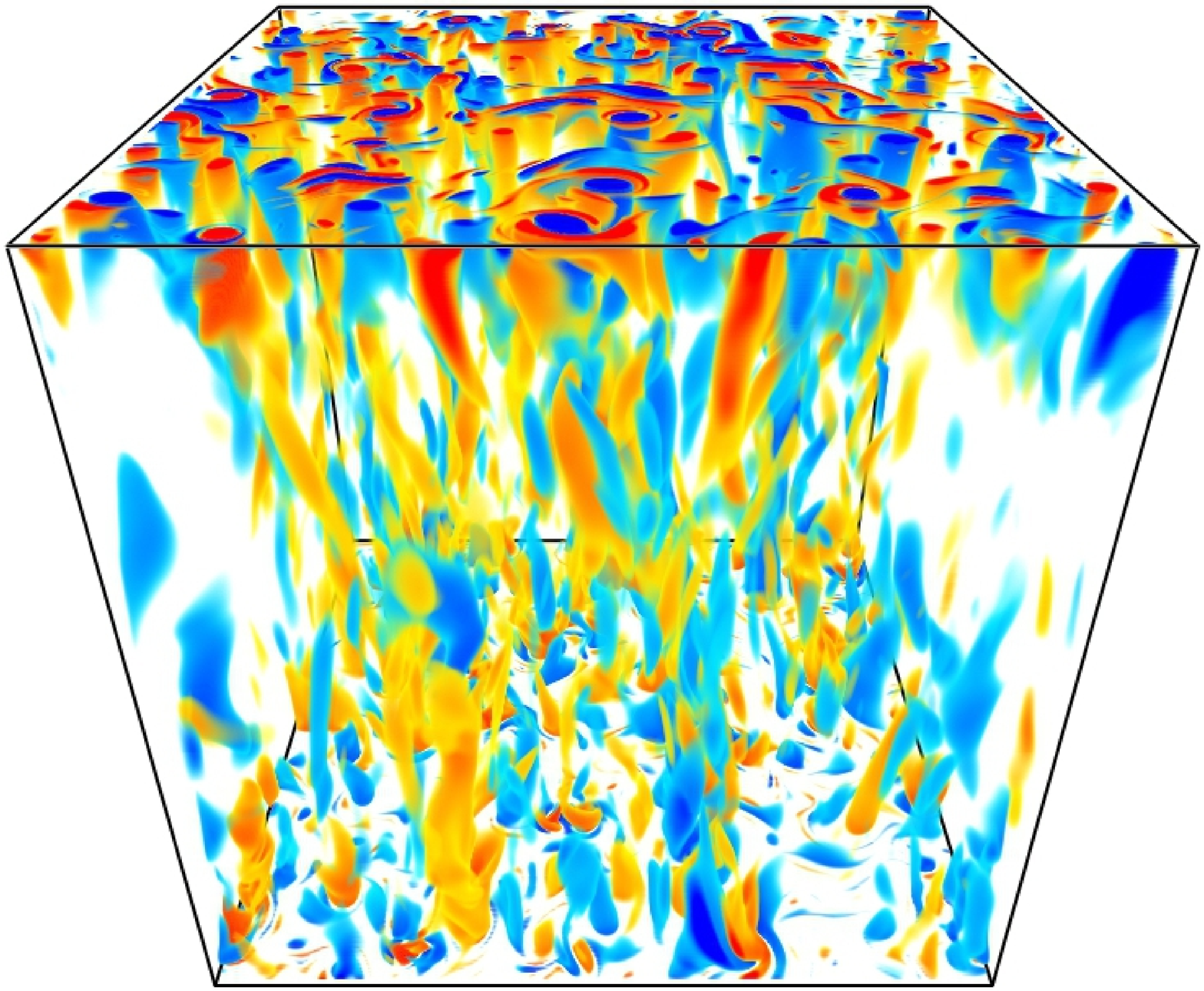}}
      \quad
    \subfloat[]{
      \includegraphics[height=4cm]{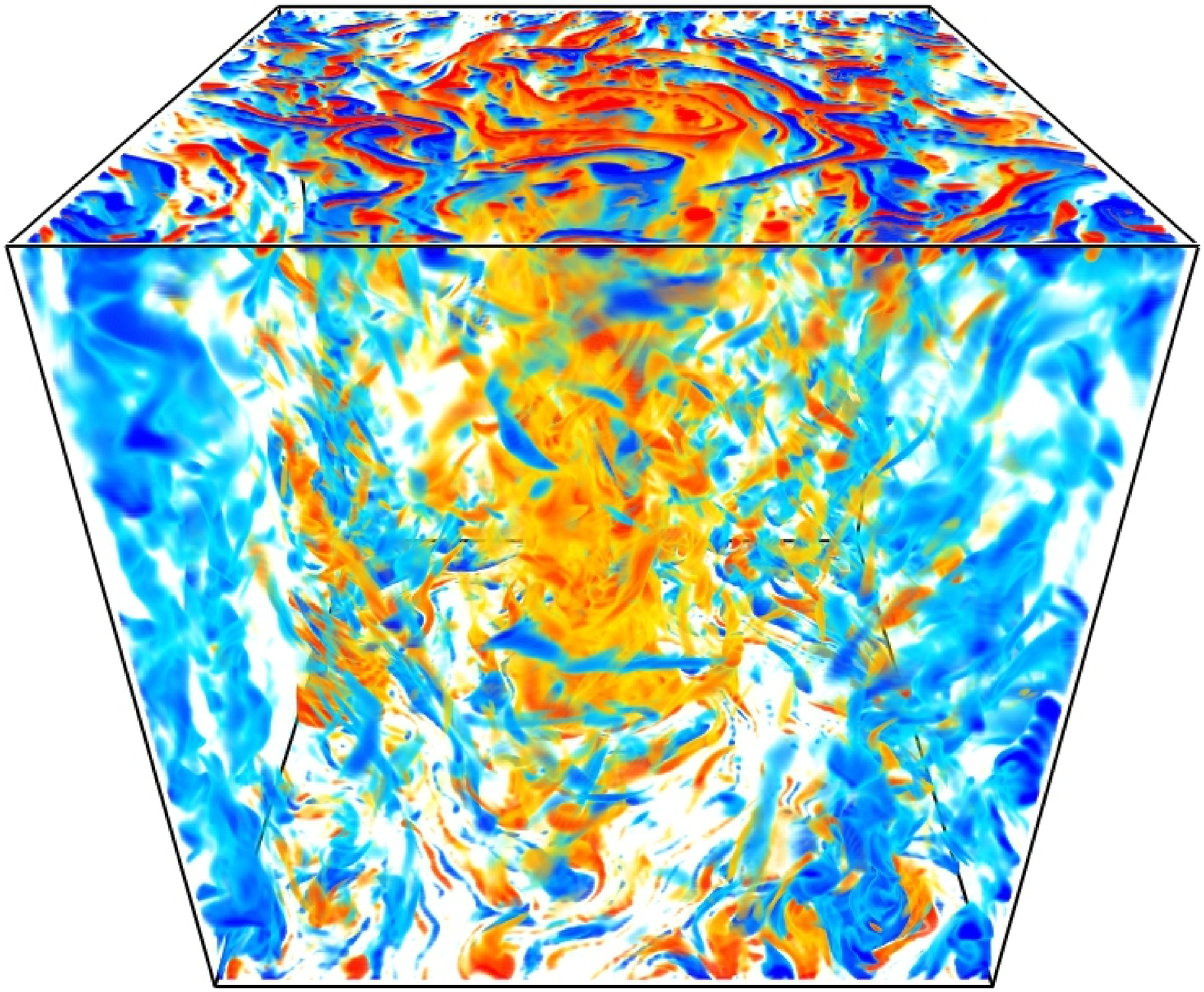}}                                 
  \end{center}
\caption{Volumetric renderings of the vertical component of the small-scale vorticity $\zeta'$ illustrating each of the four flow regimes observed in the convection simulations. (a) Cellular regime: $Pr=1$, $\Rat=10$; (b) convective Taylor column regime: $Pr=10$, $\Rat=60$; (c) plume regime: $Pr=10$, $\Rat=200$; (d) geostrophic turbulence regime: $Pr=1$, $\Rat=100$.}
\label{F:vortvol}
\end{figure}

\begin{figure}
  \begin{center}
      \includegraphics[height=7cm]{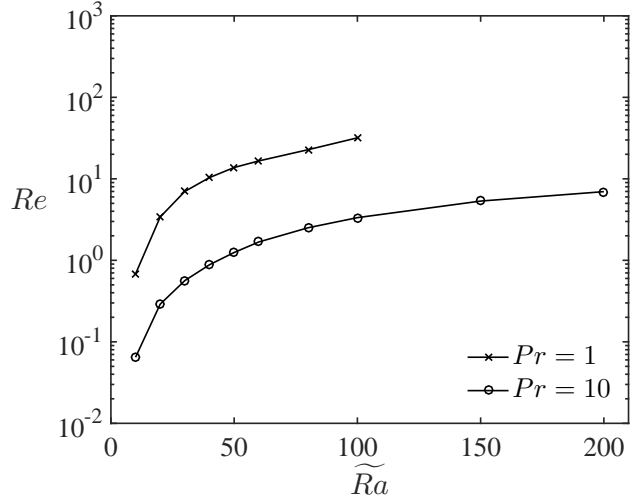}                        
  \end{center}
\caption{Convective-scale Reynolds number for all of the simulations. $Re$ is calculated from the vertically averaged root-mean-square (rms) vertical velocities.}
\label{F:Rerms}
\end{figure}

It is helpful to review some of the important characteristics of the convection that are pertinent to the dynamo problem. Figure \ref{F:vortvol} illustrates each of the four flow regimes with volumetric renderings of the axial vorticity $\zeta'$. The cellular regime  shown in Figure \ref{F:vortvol}(a) is distinguished by the cellular horizontal structure of the flow and the relative unimportance of inertia and thermal advection. The convective Taylor column (CTC) regime is exemplified by Figure \ref{F:vortvol}(b) in which the flow is characterized by sparsely populated, axially coherent structures. The plume regime of Figure \ref{F:vortvol}(c) occurs as the Rayleigh number is increased further and the CTCs become unstable and lose axial alignment. For $Pr=1$ the flow transitions to a state of geostrophic turbulence near $\Rat \approx 50$, and the strongly turbulent case shown in Figure \ref{F:vortvol}(d) is representative of the flow morphology in this regime. For a detailed report of flow morphology, statistics, and balances in rapidly rotating convection we refer the reader to \citet{kJ12}.

In Figure \ref{F:Rerms} we show the convective-scale Reynolds number as a function of $\Rat$ for all of the cases investigated. Note that in the rapidly rotating, quasi-geostrophic limit the Reynolds number based on the depth of the fluid layer is given by $Re_H = \ep^{-1} Re$, and thus intrinsically large for $Re=O(1)$. The plotted values of $Re$ were obtained by taking vertical averages of the root-mean-square (rms) vertical velocities, where the mean is defined as an average over the horizontal plane and time. The Reynolds numbers for the different Prandtl number cases are separated approximately by an order of magnitude for a given value of $\Rat$, and the data points for both cases show the same general trend with increasing $\Rat$. For $Pr=1$, the turbulent regime ($\Rat \gtrsim 50$) is characterized by $Re \gtrsim 13$. For $Pr=10$ the highest Reynolds number simulation carried out was for $\Rat = 200$ where $Re \approx 7$; in comparison, a similar value of the Reynolds number is observed for $Pr=1$ at a much lower Rayleigh number of $\Rat = 30$.

\begin{figure}
  \begin{center}
   \subfloat[]{
      \includegraphics[height=4.5cm]{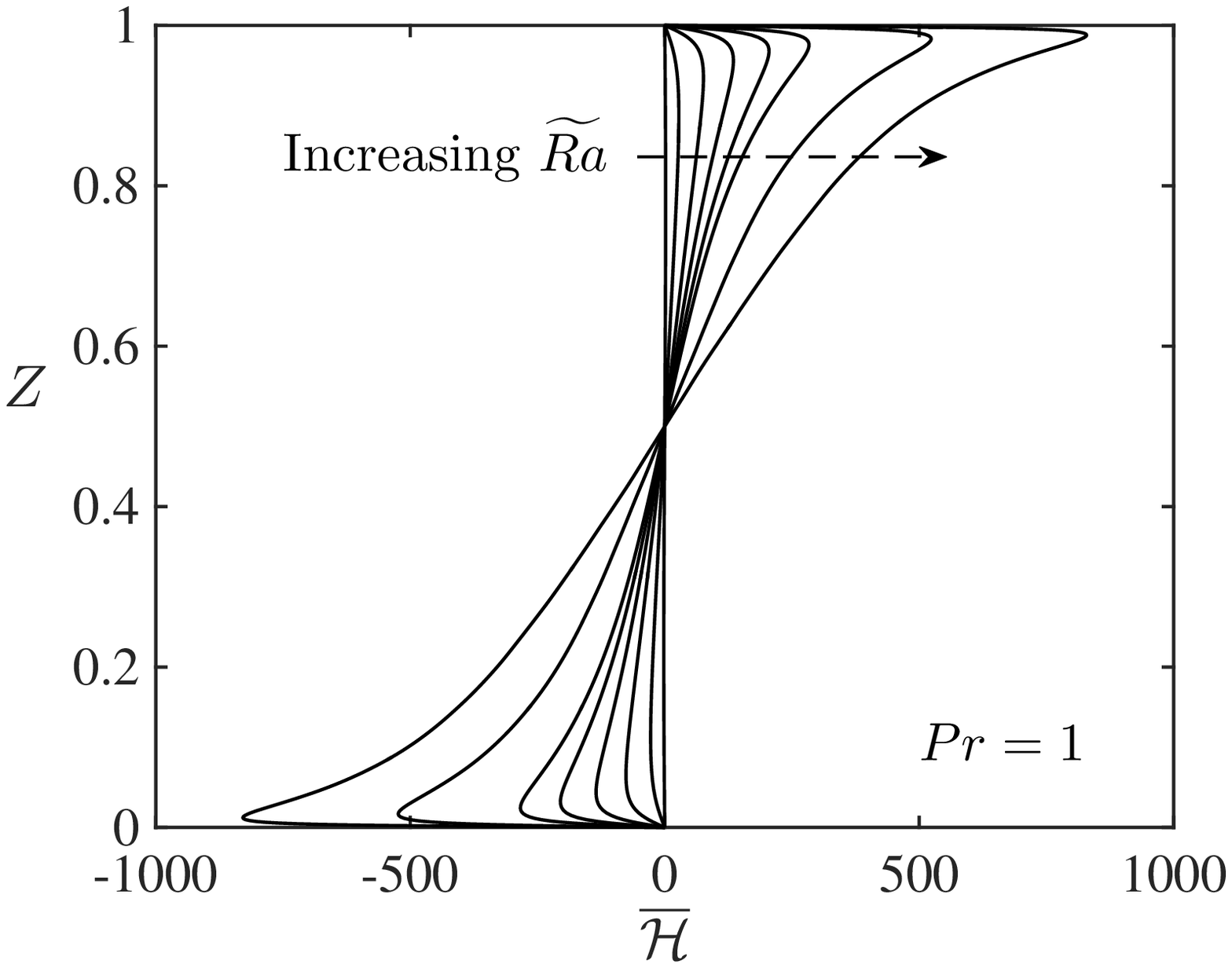}}
      \qquad
    \subfloat[]{
      \includegraphics[height=4.5cm]{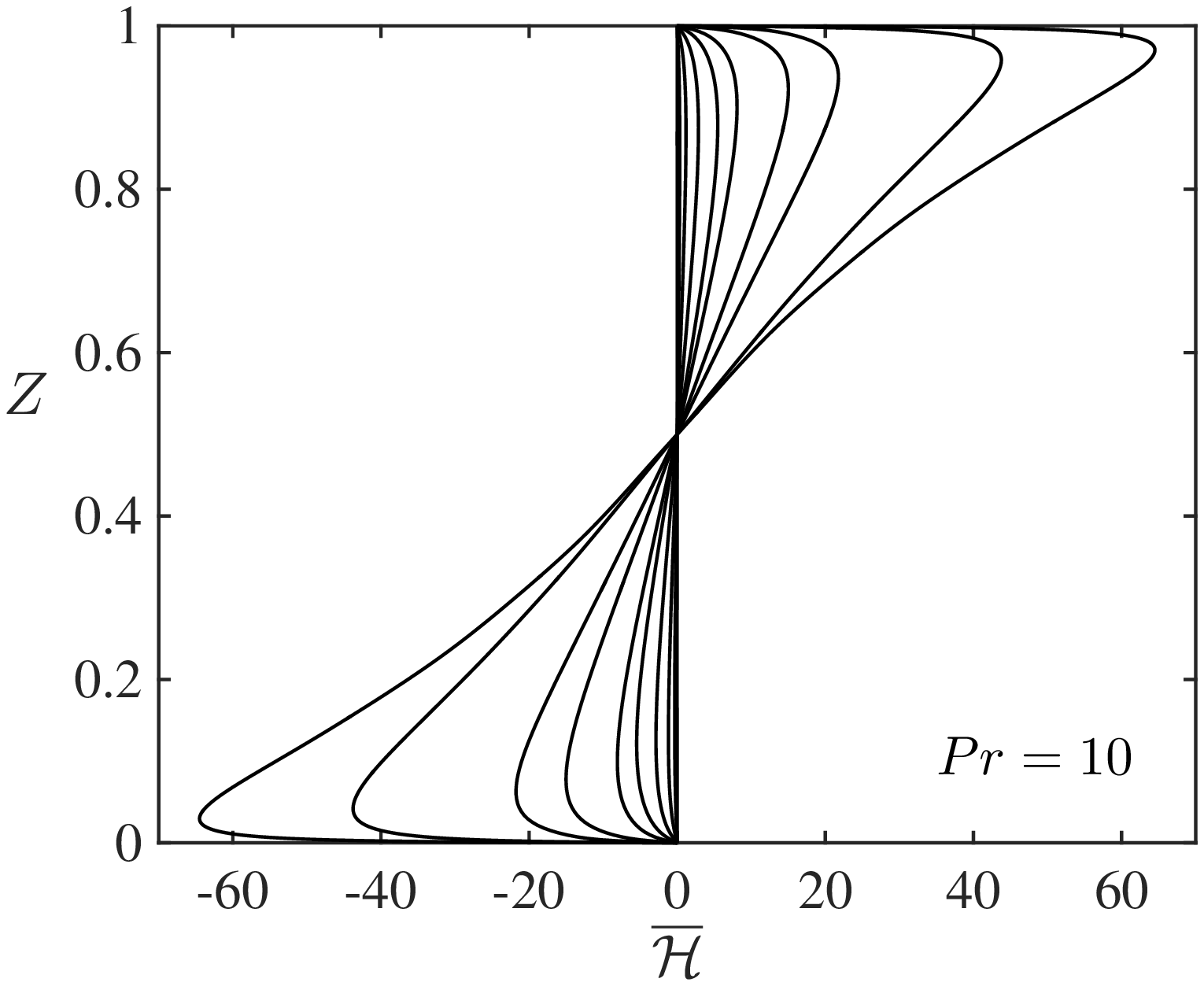}}  \\ 
   \subfloat[]{
      \includegraphics[height=4.5cm]{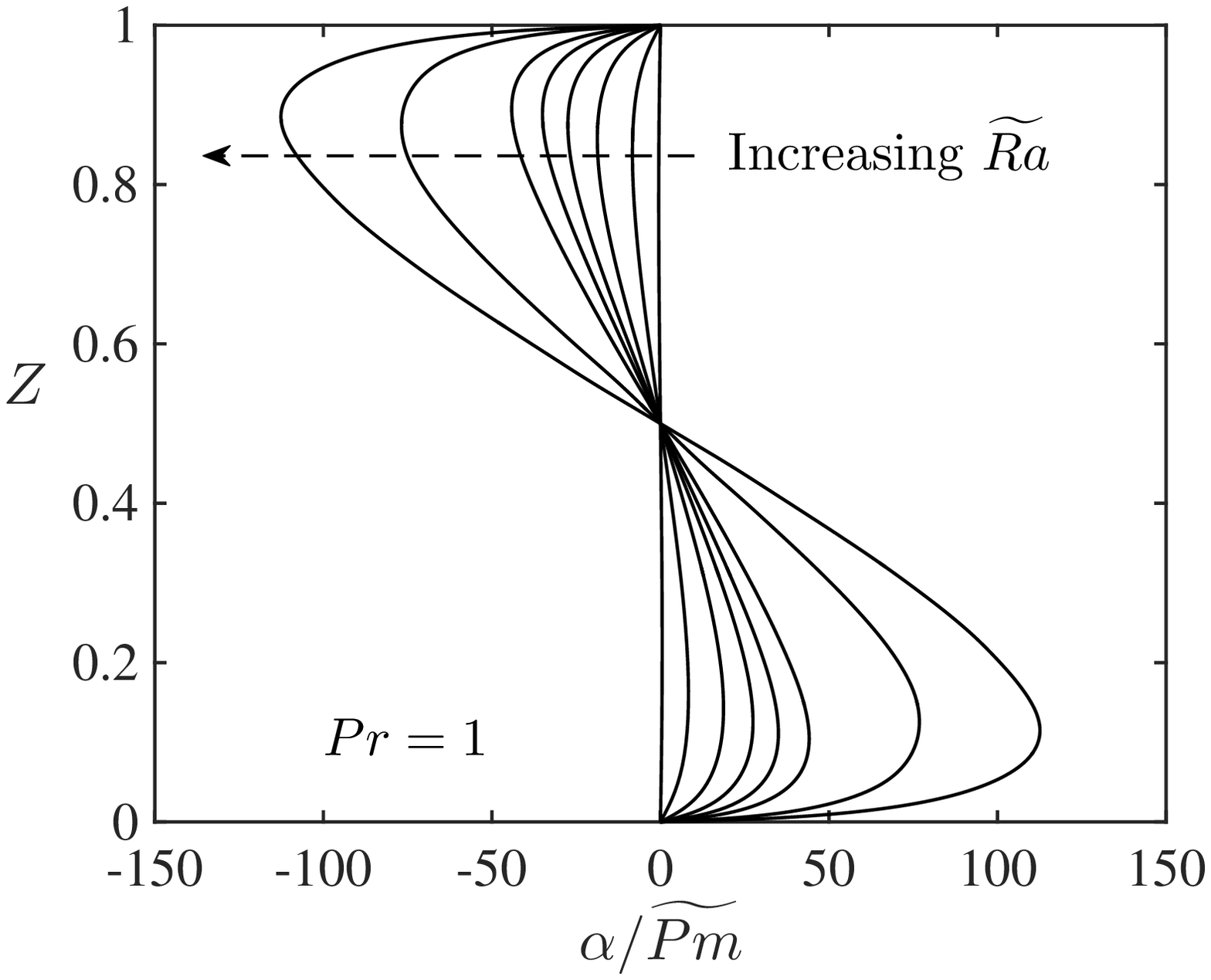}}
      \qquad
    \subfloat[]{
      \includegraphics[height=4.5cm]{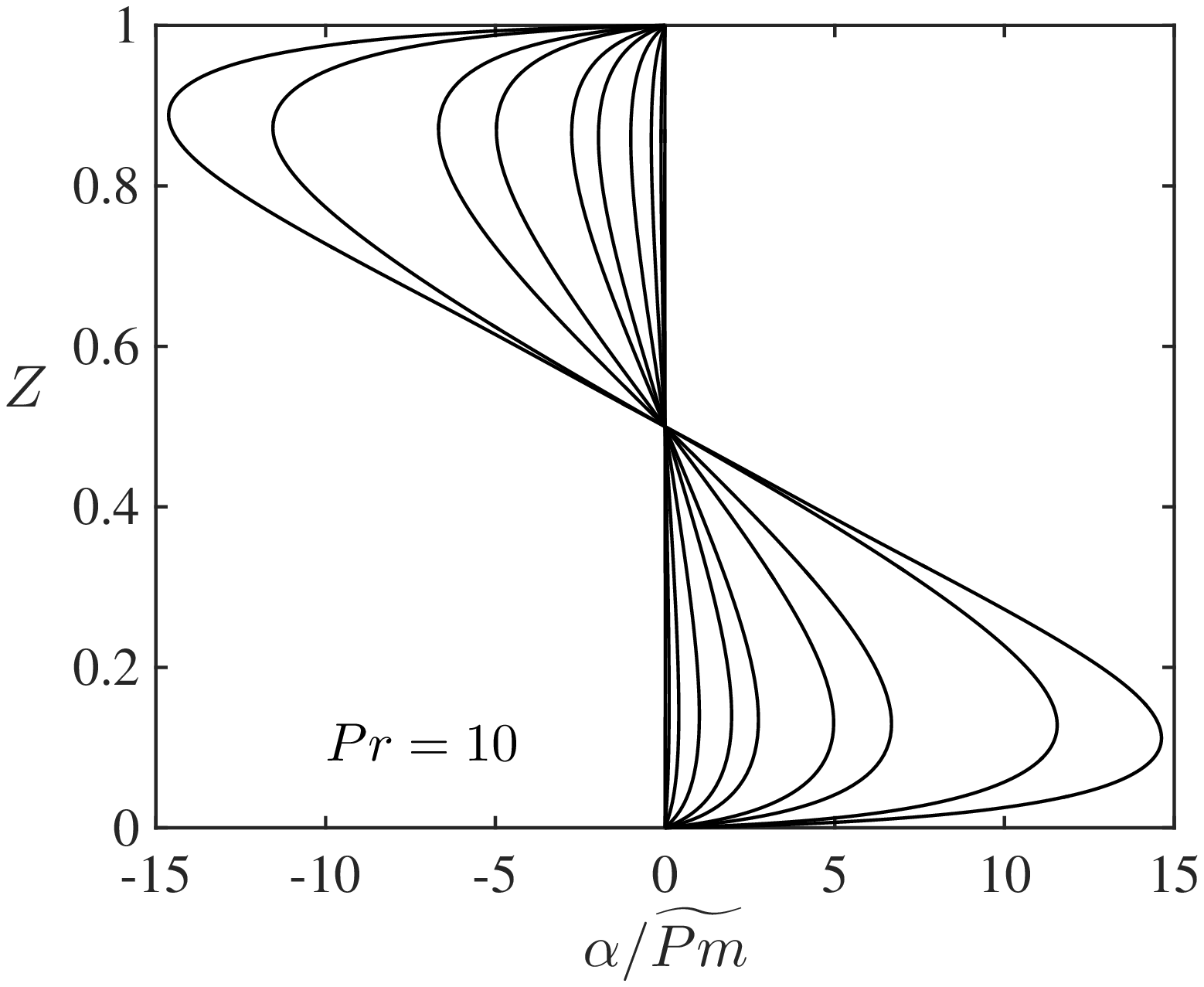}}                        
  \end{center}
\caption{Vertical profiles of helicity for (a) $Pr=1$ and (b) $Pr=10$; and vertical profiles of $\alpha/\Pmt$ for (c) $Pr=1$ and (d) $Pr=10$ for all investigated Rayleigh numbers.}
\label{F:hel}
\end{figure}

Profiles of $\He$ and $\alpha$ are shown in Figure \eqref{F:hel} for all of the simulations. We do not find a significant change in the shape of $\alpha$ as $\Rat$ is increased, though the location of the maximal values does exhibit a shift towards the boundaries of the domain with increasing $\Rat$. The $\He$ profiles show more obvious boundary layer behavior for both Prandtl numbers; these profiles are reminiscent of the maximum Nusselt number single mode solutions investigated previously for the QGDM \citep{mC16}. In contrast, the $\alpha$ profiles are more reminiscent of the fixed-wavenumber single mode solutions which become self-similar with increasing $\Rat$ \citep[e.g.~see Figure 4 of][]{mC16}. The $\He$ boundary layer is more pronounced for $Pr=1$ where higher Reynolds numbers are accessed.

\begin{figure}
  \begin{center}
      \includegraphics[height=5cm]{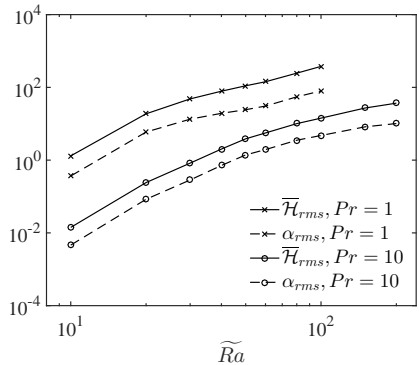}
  \end{center}
\caption{Vertical rms values of $\He$ and $\alpha$ as a function of $\Rat$ for the two different Prandtl numbers.}
\label{F:helrms}
\end{figure}

The rms values of $\He$ and $\alpha$ are given in Figure \ref{F:helrms} and show that both functions have nearly identical scaling behavior with increasing Rayleigh number, though we find that $\He$ increases slightly more rapidly when viewed on a linear abscissa scale. As for the Reynolds number based on the rms convective velocity shown in Figure \ref{F:Rerms}, the influence of the Prandtl number appears only to influence the magnitude of the rms values for each function, but the general qualitative increase with $\Rat$ is found for both Prandtl numbers.

\begin{figure}
  \begin{center}
   \subfloat[]{
      \includegraphics[height=5cm]{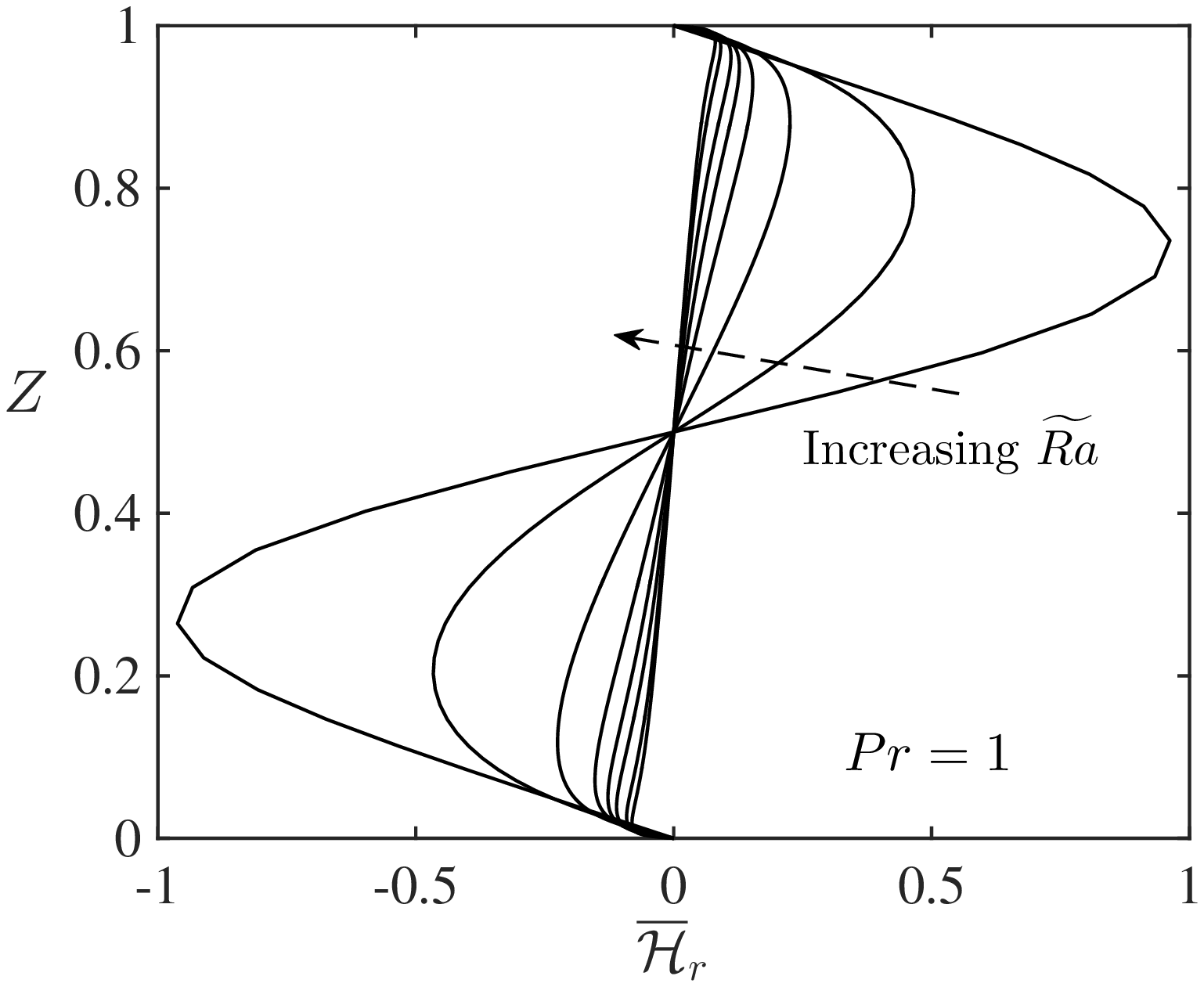}}
      \qquad
    \subfloat[]{
      \includegraphics[height=5cm]{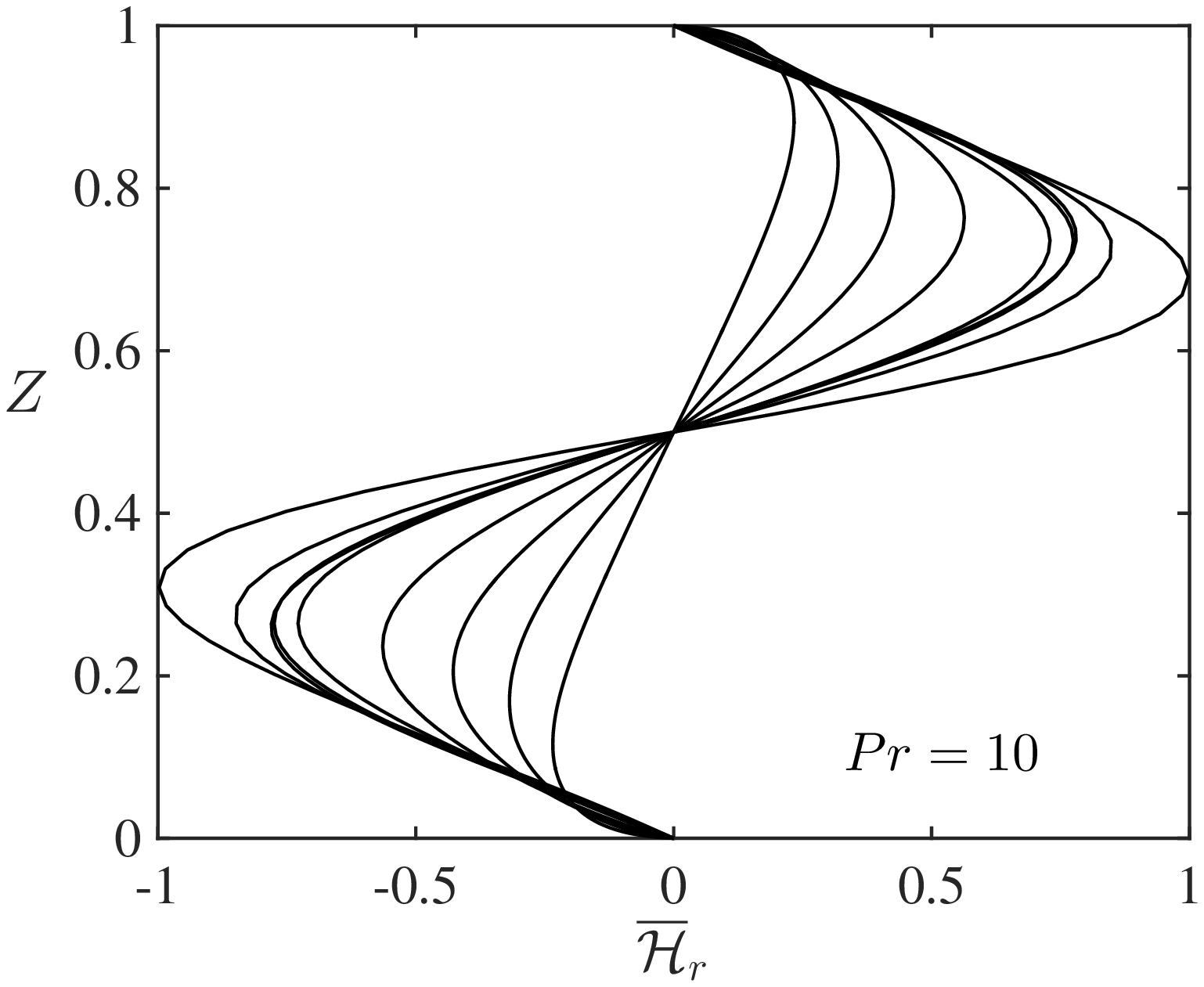}}                        
  \end{center}
\caption{Profiles of the relative helicity $\He_r$ for (a) $Pr=1$ and (b) $Pr=10$ for all of the investigated $\Rat$; higher values of $\Rat$ have smaller overall values of $\He_r$ (see Figure \ref{F:relhelrms}).}
\label{F:relhel}
\end{figure}

Figure \ref{F:relhel} shows profiles of the relative helicity $\He_r$. For both Prandtl numbers the general trend observed is an overall decrease in the magnitude of $\He_r$ and a shift towards the boundaries of the maximum value of $\He_r$; these results are in excellent agreement with the DNS study of \citet{sST10}. For $Pr=10$ we find that the two profiles for $\Rat=30$ and $\Rat=40$ possess nearly identical $\He_r$ profiles despite the fact that the magnitudes of both $\alpha$ and $\He$ are different. This situtation will arise if the solutions have the same spatial form, but differing amplitudes.

\begin{figure}
  \begin{center}
      \includegraphics[height=6cm]{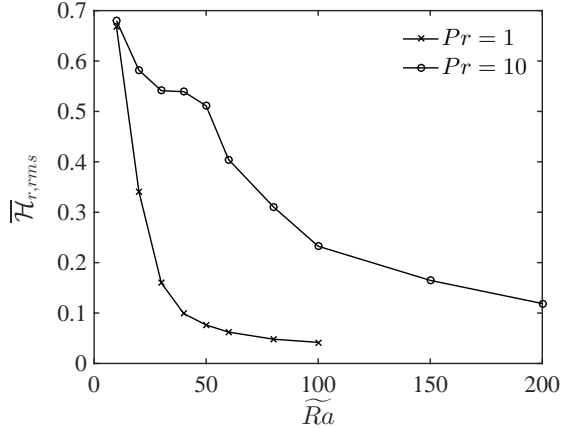}
   \end{center}
\caption{Vertical rms values of the relative helicity, $\He_{r,rms}$, as a function of $\Rat$ for all of the cases investigated.}
\label{F:relhelrms}
\end{figure}

Vertical rms values of the relative helicity $\He_{r,rms}$ are given in Figure \ref{F:relhelrms}. Given the results of Figure \ref{F:relhel} it is unsurprising to see the general decrease of $\He_{r,rms}$ with increasing $\Rat$; this result is again in excellent agreement with \citet{sST10} and the spherical investigation of \citet{kS12} where the same behavior was observed. We find that $\He_{r,rms}$ decreases more rapidly for $Pr=1$ than for $Pr=10$; the same Prandtl number trend was also observed in \citet{sST10}. The more rapid $\He_{r,rms}$ decrease observed for $Pr=1$ is due likely to the enhanced role of inertia and more irregular flows. Given that the reduced system of equations models convection only in the low Rossby number regime, we can say that the overall decrease in $\He_{r,rms}$ need not be associated with a loss in rotational constraint. Rather, the decrease of $\He_{r,rms}$ is due to the change in convective flow behavior to more disordered states where the maximally helical nature of convection near onset ($\Rat \approx \Rat_c$) is lost. Nevertheless, it should be recalled that the rms values of both $\He$ and $\alpha$ still show a monotonic increase with $\Rat$ (see Figure \ref{F:helrms}), though the rate of growth appears to slow with increasing $\Rat$.

\begin{figure}
  \begin{center}
   \subfloat[]{
      \includegraphics[height=5.1cm]{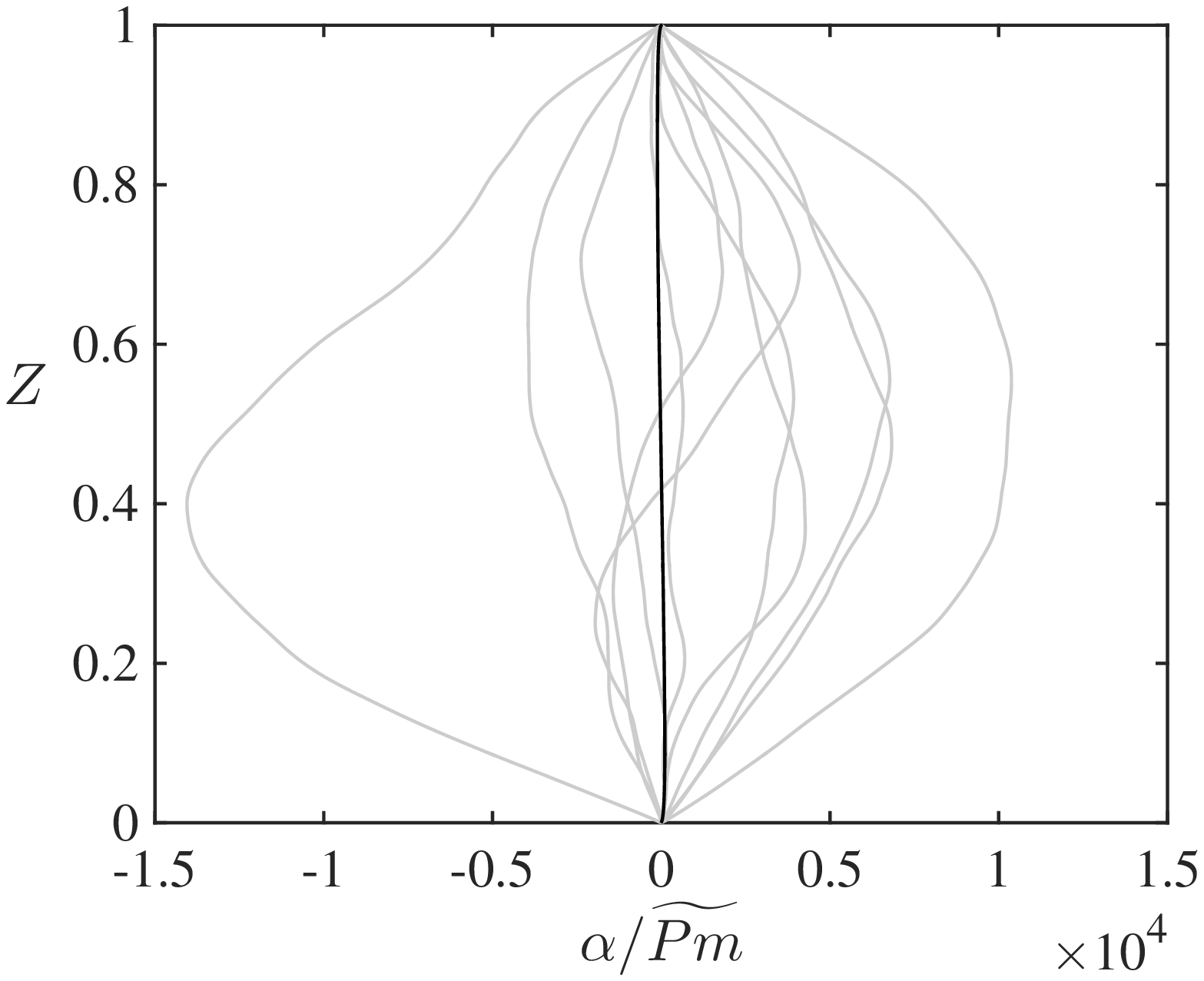}}
      \qquad
    \subfloat[]{
      \includegraphics[height=5.1cm]{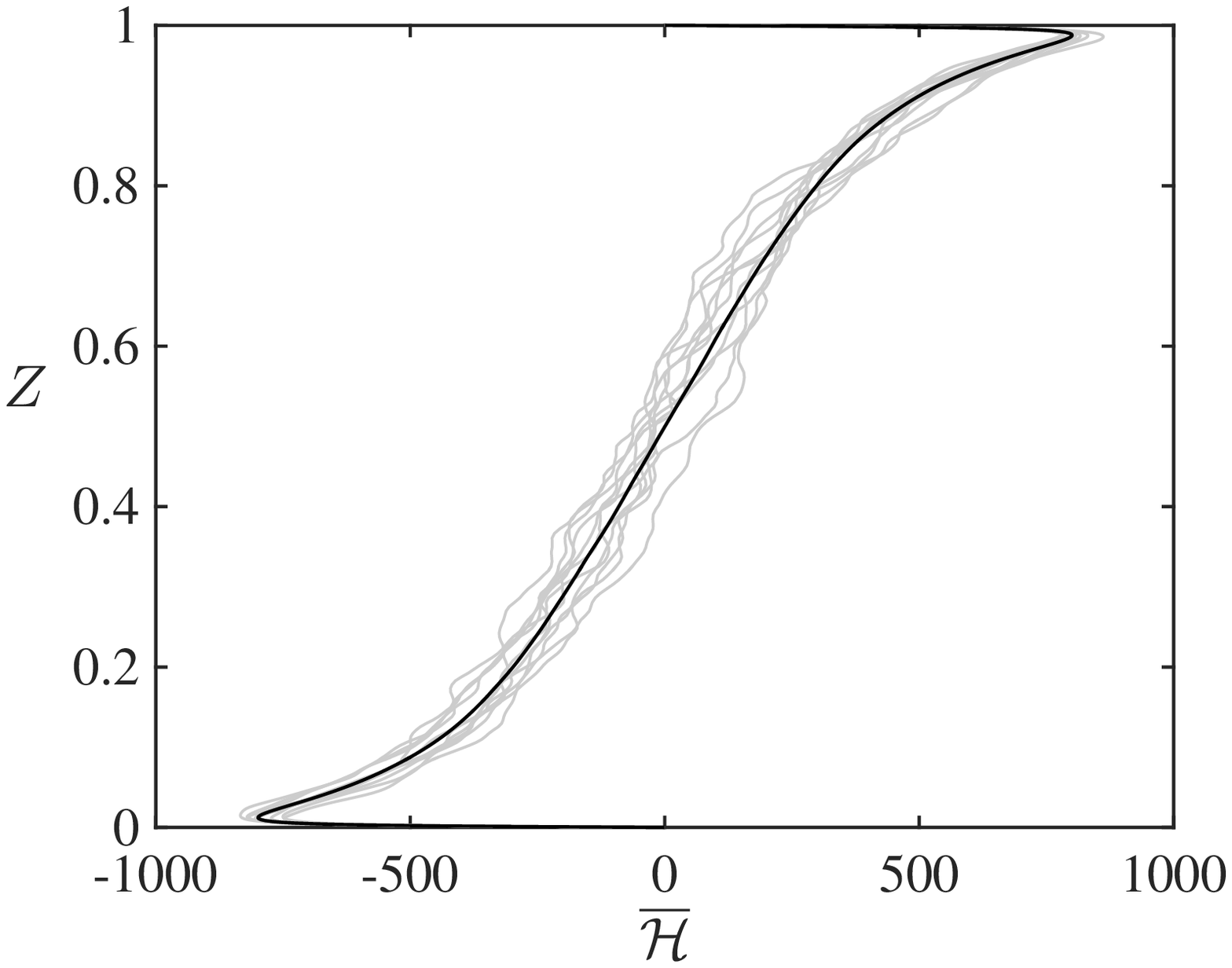}}                           
  \end{center}
\caption{Comparison of the instantaneous, horizontally averaged (gray) and time and horizontally averaged (black) (a) $\alpha/\Pmt$ and (b) $\He$ profiles for $Pr=1$ and $\Rat=100$. Note the significant variation in the instantaneous (gray) $\alpha$ profiles relative to the time-averaged (black) profiles in (a); significantly less temporal variation is observed for $\He$ in (b).}
\label{F:alphinst}
\end{figure}

The simulations show that $\alpha$ is characterized by more pronounced temporal variations in comparison to $\He$, which converges much more rapidly over the course of a simulation. This is similar behaviour to that observed by \citet{fC06} who compared the statistical properties of $\alpha$ in relation to those for $\He$ for moderately rotating turbulent convection. In Figure \ref{F:alphinst} we have plotted both the instantaneous, horizontally averaged (in gray) profiles and the time and horizontally averaged profiles (in black). One of the most obvious differences between $\alpha$ and $\He$ is the extreme temporal variability shown in the former quantity where large variations in the profile are observed. For $\He$ the instantaneous profiles possess the same general characteristics as the time-averaged profile, whereas the same cannot be said of $\alpha$. This difference in behavior may be due to the presence of the inverse Laplacian in the definition of $\alpha$ (see equation \eqref{E:alpha}) since this operator tends to enhance low wavenumber, large-scale structures that possess significant kinetic energy and large temporal fluctuations.

\subsection{Kinematic dynamos}

\begin{figure}
  \begin{center}
   \subfloat[]{
      \includegraphics[height=5cm]{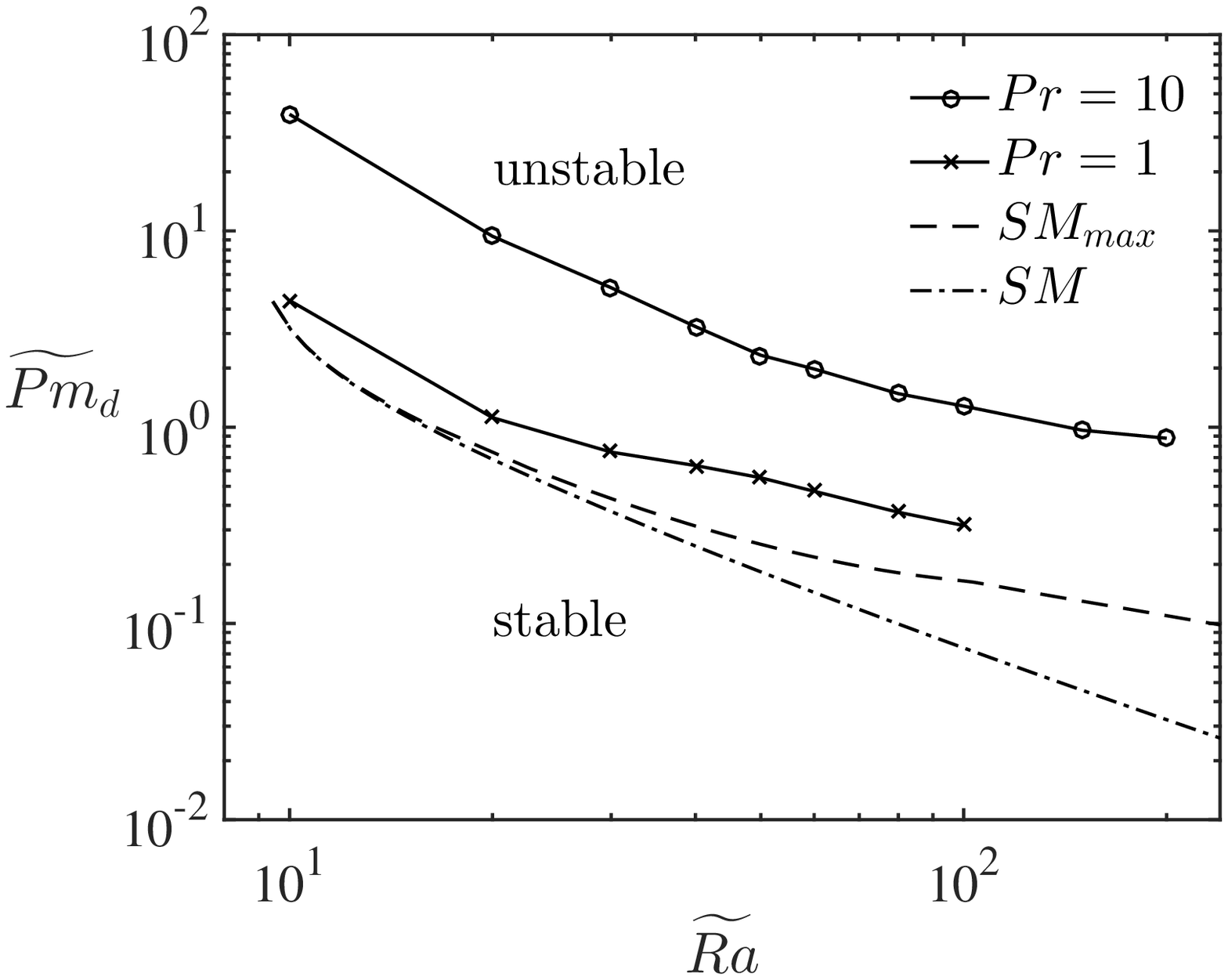}}
      \qquad
    \subfloat[]{
      \includegraphics[height=5cm]{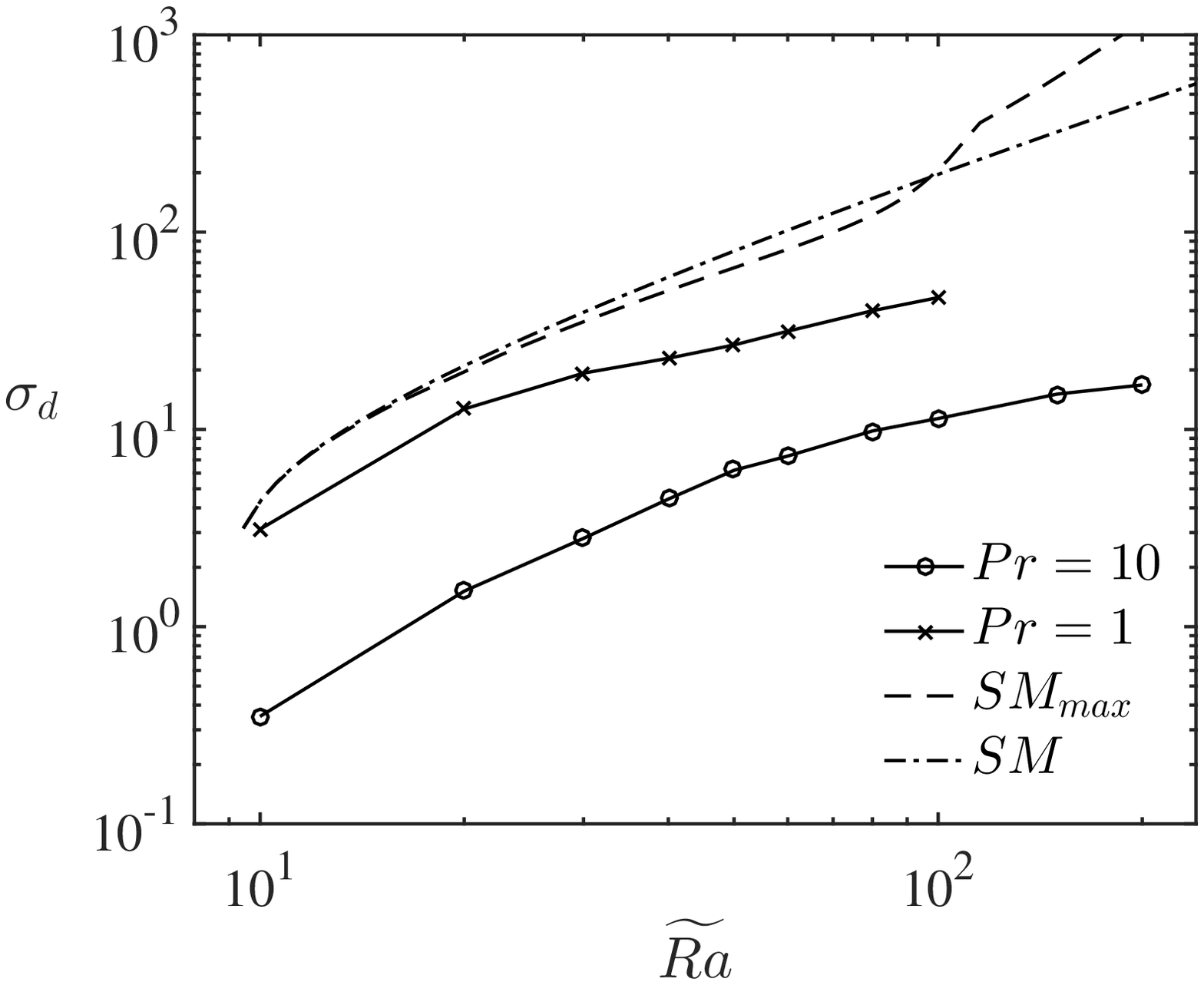}}                        
  \end{center}
\caption{Kinematic dynamo results showing (a) the critical magnetic Prandtl number $\Pmt_d$ and (b) the critical dynamo frequency $\sigma_d$ versus $\Rat$. The steady single mode solutions of \citet{mC16} are shown by the broken curves: $SM_{max}$ are those solutions that maximize the heat transfer and have wavenumbers $\kt$ which vary (increase) with increasing $\Rat$; solutions $SM$ have a fixed horizontal wavenumber of $\kt=\kt_c=1.3048$ for all $\Rat$.}
\label{F:kin}
\end{figure}

For a given value of $\Rat$, we compute the minimum value of $\Pmt$ that yields a mean magnetic field with zero growth rate (marginal stability); the corresponding critical value of the magnetic Prandtl number is denoted as $\Pmt_d$. As in \citet{mC16}, all of the dynamos are found to be oscillatory with a critical dynamo frequency $\sigma_d$. Figures \ref{F:kin}a and b show $\Pmt_d$ and $\sigma_d$ versus $\Rat$. The results from the steady single mode solutions of \citet{mC16} are also shown: $SM_{max}$ are those solutions that maximize the heat transfer and have wavenumbers $\kt$ which vary (increase) with increasing $\Rat$; solutions $SM$ have a fixed horizontal wavenumber of $\kt=\kt_c=1.3048$ for all $\Rat$. Despite the observed changes in flow regime over the investigated range of $\Rat$, we find little evidence for these changes in the dynamo behavior; for steady single mode solutions these results are insensitive to $Pr$. This observation might have been anticipated given the similarities found in the $\alpha$ profiles for all $\Rat$ of Figure \ref{F:hel}. For instance, for $Pr=1$ the flow transitions to turbulence near $\Rat =50$, yet the $\Pmt_d$ and $\sigma_d$ data do not show any pronounced changes in this region of parameter space. This result might seem surprising given that such transitions in flow regime are characterized by distinct statistics \citep{kJ12,dN14}. 

In general, the results show that the magnitude of the $\Pmt_d-\Rat$ slope is a decreasing function of $\Rat$, resulting in a general decrease in the efficiency of the convection-driven dynamo. It's possible that this behavior is linked to the rapid decrease in $\He_r$ observed in Figures \ref{F:relhel}(a) and \ref{F:relhelrms}.  Figure \ref{F:kin}(b) shows that lower values of $\Pmt$ are also associated with higher values of the dynamo frequency $\sigma_d$, since higher values of $Re$ are required.

\begin{figure}
  \begin{center}
   \subfloat[]{
      \includegraphics[height=5cm]{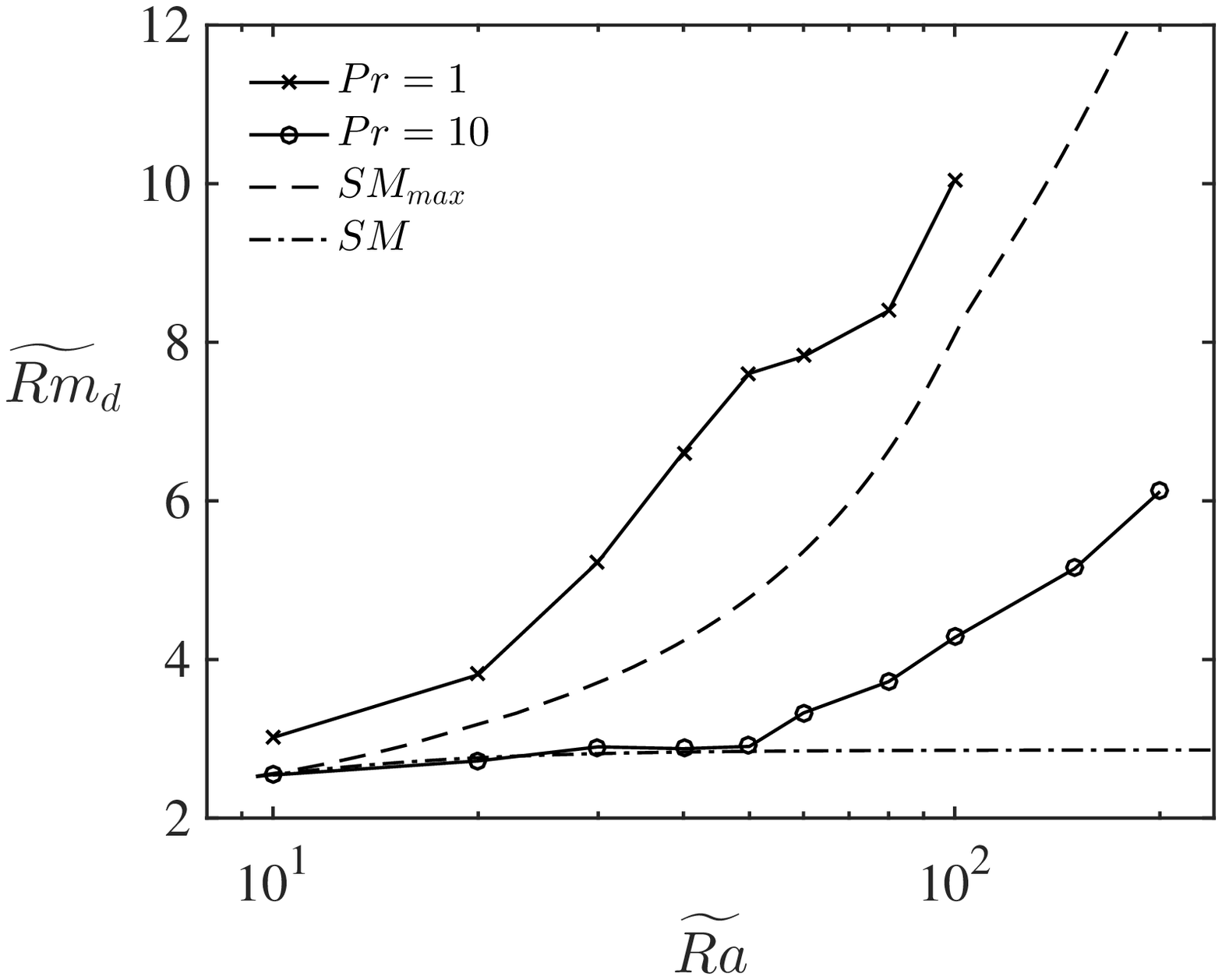}}
      \qquad
    \subfloat[]{
      \includegraphics[height=5cm]{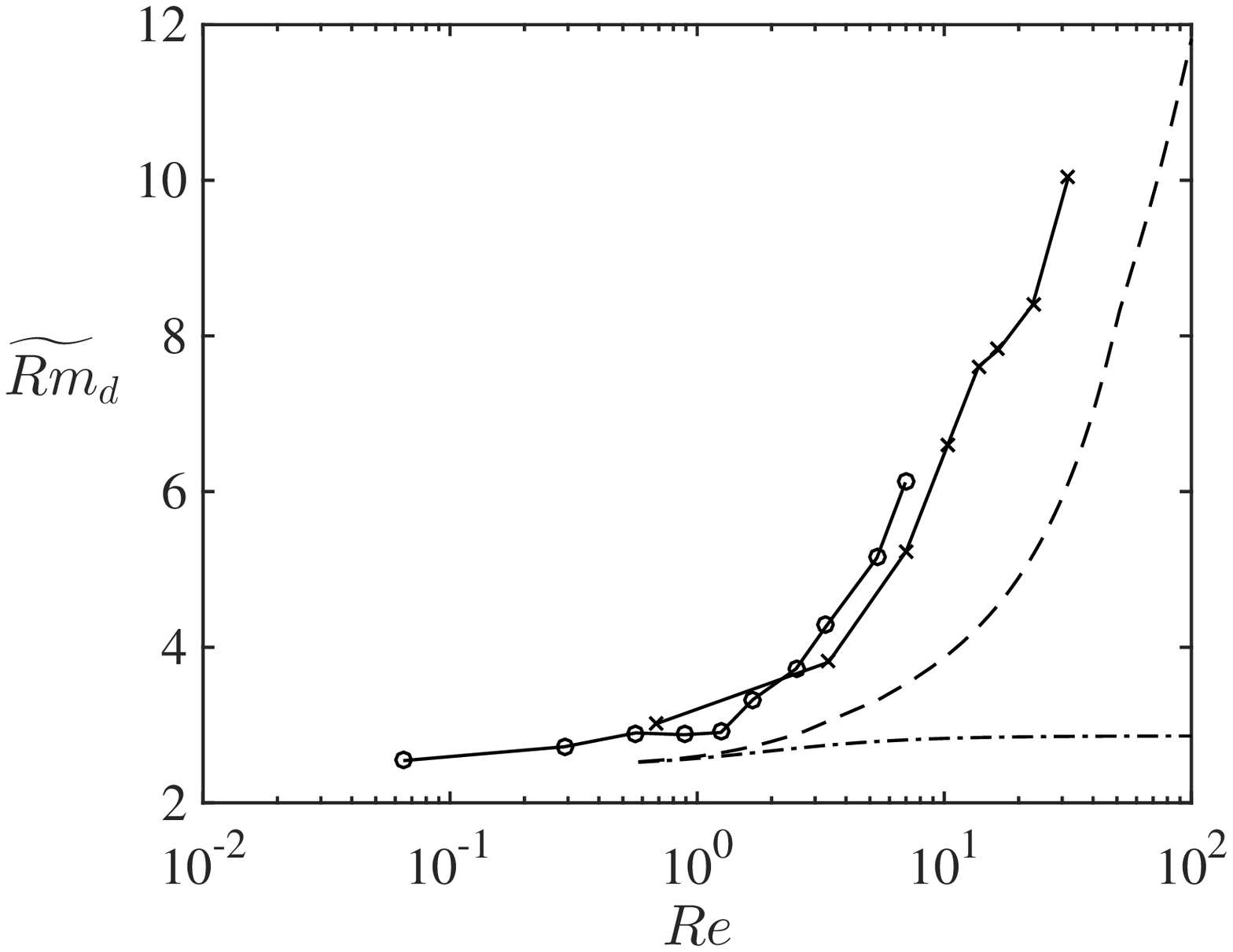}}                       
  \end{center}
\caption{Critical reduced magnetic Reynolds number $\Rmt_d$ versus (a) $\Rat$ and (b) $Re$ for all of the cases investigated. The steady single mode solutions of \citet{mC16} are shown by the broken curves.}
\label{F:RmRe}
\end{figure}

The critical reduced magnetic Reynolds number $\Rmt_d = \Pmt_d Re$ is shown as a function of $\Rat$ and $Re$ in Figures \ref{F:RmRe}a and b, respectively. In general, we find that $\Rmt_d$ increases with $\Rat$ (and $Re$), with the most significant increases observed for the $Pr=1$ cases. The $Pr=10$ cases show a similar trend as the $Pr=1$ cases when viewed from the perspective of Figure \ref{F:RmRe}(b). The $Pr=10$ cases closely follow the $SM$ solutions up to $\Rat=50$, where the curve then shows a more substantial increase in $\Rmt_d$; this is the region in parameter space where the flow begins to transition from the CTC regime to the plume regime \citep{kJ12,dN14}. We also find a change in behavior at $\Rat=50$ for the $Pr=1$ simulations where the flow transitions to turbulence, and the rate of increase in $\Rmt_d$ slows briefly before picking up again at $\Rat=100$. The $SM_{max}$ solutions do exhibit the same general trend of increasing $\Rmt_d$ with $\Rat$ and $Re$ as the numerical simulations, in contrast to the fixed wavenumber $SM$ solutions which asymptote to a constant value of $\Rmt_d \approx 2.9$ as $(\Rat,Re) \rightarrow \infty$. In contrast to the $\Pmt_d$ results shown in Figure \ref{F:kin}, here we find that the $\Rmt_d$ behavior is, at least to some extent, reflective of flow regime behavior.

\begin{figure}
  \begin{center}
   \subfloat[]{
      \includegraphics[height=5cm]{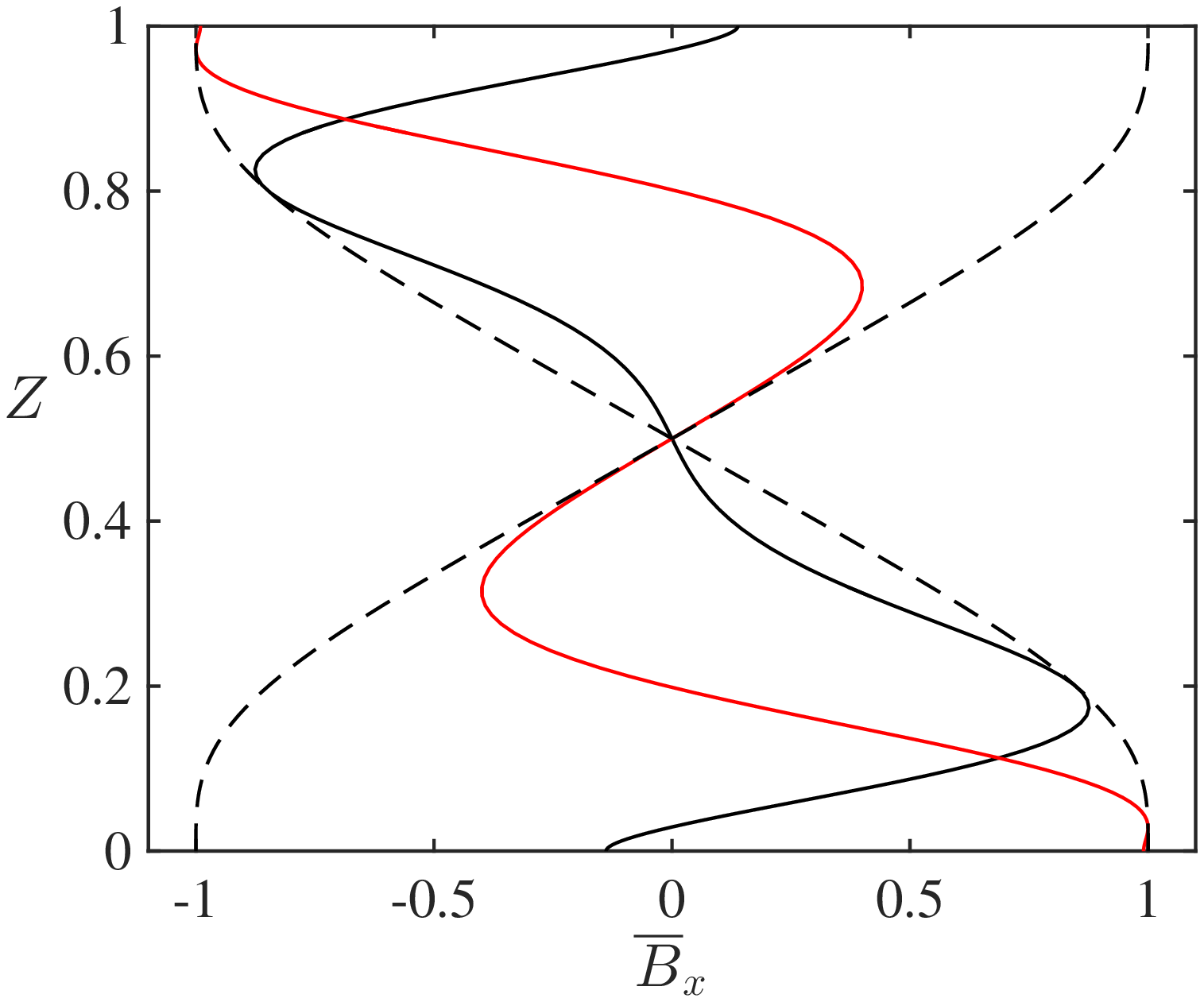}}
      \qquad       
   \subfloat[]{
      \includegraphics[height=5cm]{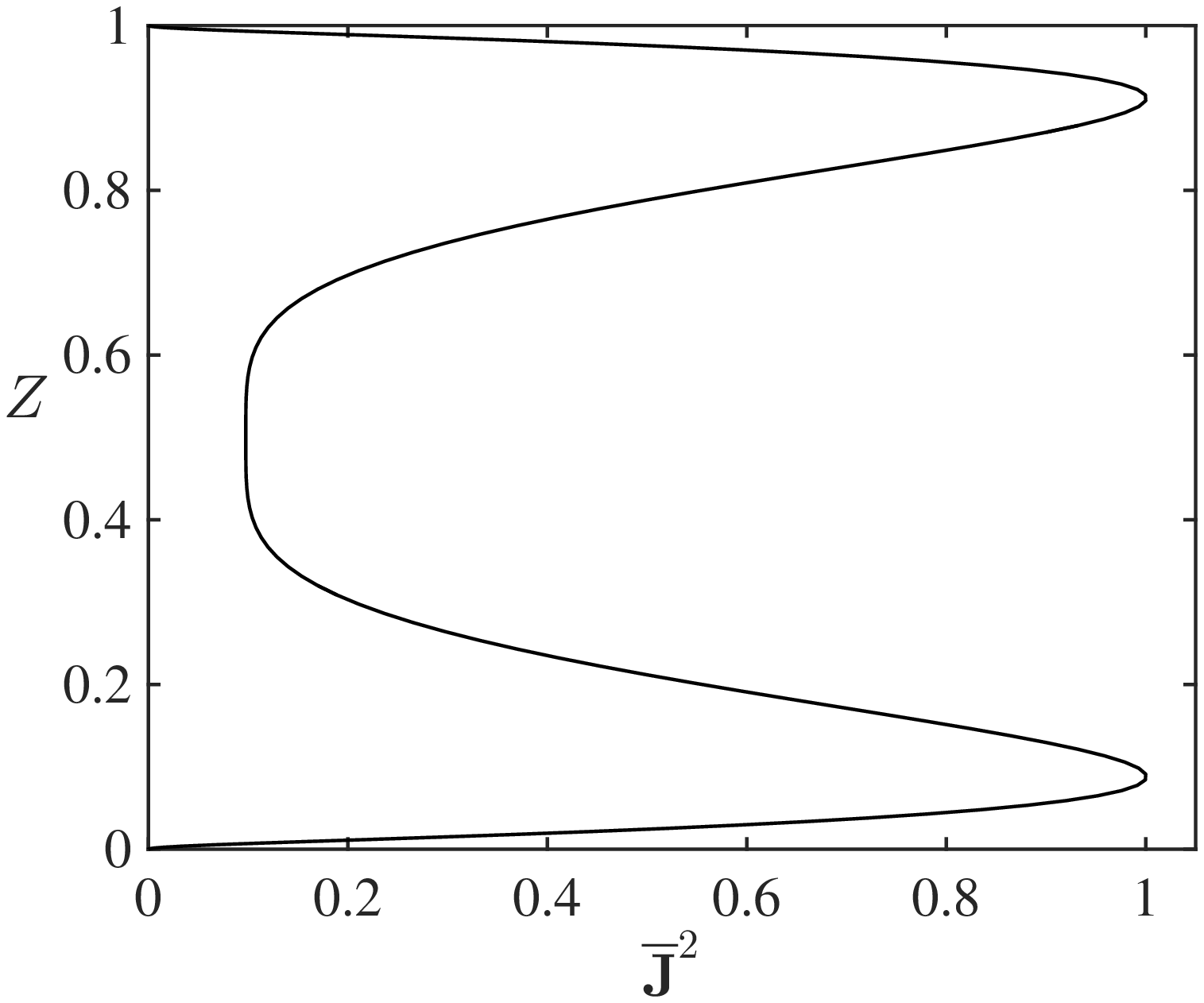}}                
  \end{center}
\caption{(a) $x$-component of the marginally stable mean magnetic field eigenfunction $\mBx$ for $Pr=1$ and $\Rat=50$. The real and imaginary parts of the solution are shown by the solid black and solid red curves, respectively, and the modulus is given by the dashed black curves. (b) Mean ohmic dissipation $\mJb^2$ for the eigenfunction shown in (a).}
\label{F:Bprof}
\end{figure}

Profiles of the marginally stable mean magnetic field and ohmic dissipation $\mJb^2$ are given in Figure \ref{F:Bprof} for $\Rat=50$ and $Pr=1$. In accordance with the observed properties of the $\alpha$ profiles shown in Figure \ref{F:hel}, we find that the mean magnetic field structure does not change significantly as the Rayleigh and Prandtl numbers are varied; the case shown is therefore representative of all the cases investigated to a high degree of approximation. The profile shown in Figure \ref{F:Bprof}(a) is nearly identical in structure to the mean magnetic field profiles obtained from the $SM$ solutions of \citet{mC16} (see their Figure 11) and also previous weakly nonlinear \citep{kM13} and linear investigations \citep{bF13}. For the perfectly conducting boundary conditions employed here, the mean magnetic field is antisymmetric with respect to the midplane ($Z=0.5$) of the fluid layer, leading to a mean current density that is symmetric across the midplane. The ohmic dissipation $\mJb^2$ shown in Figure \ref{F:Bprof}(b) reaches maximum values where vertical gradients in $\mBb$ are largest ($Z\approx 0.1,0.9$ for the case shown).

\begin{figure}
  \begin{center}
   \subfloat[]{
      \includegraphics[height=5.5cm]{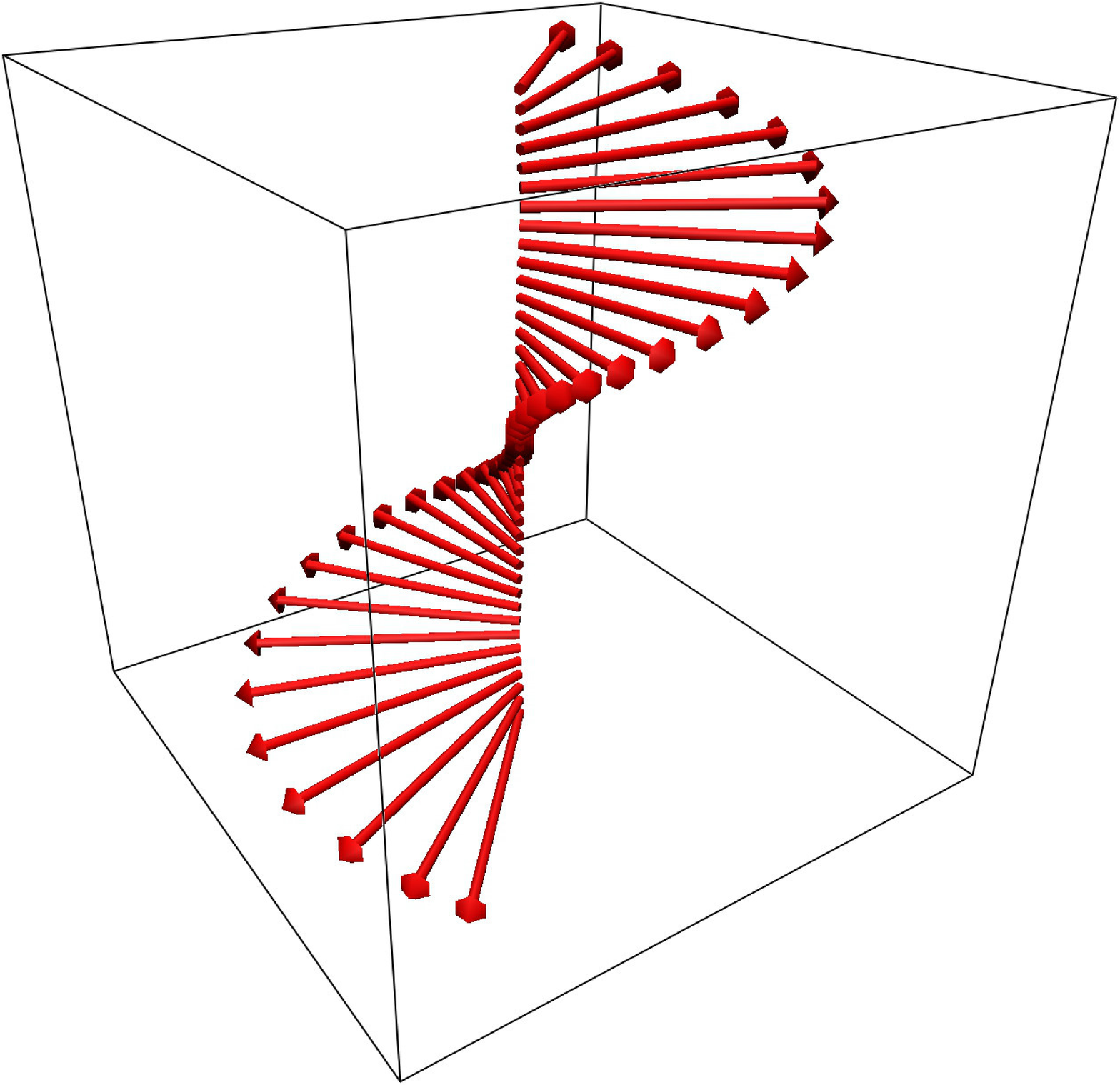}}
      \qquad
    \subfloat[]{
      \includegraphics[height=5.5cm]{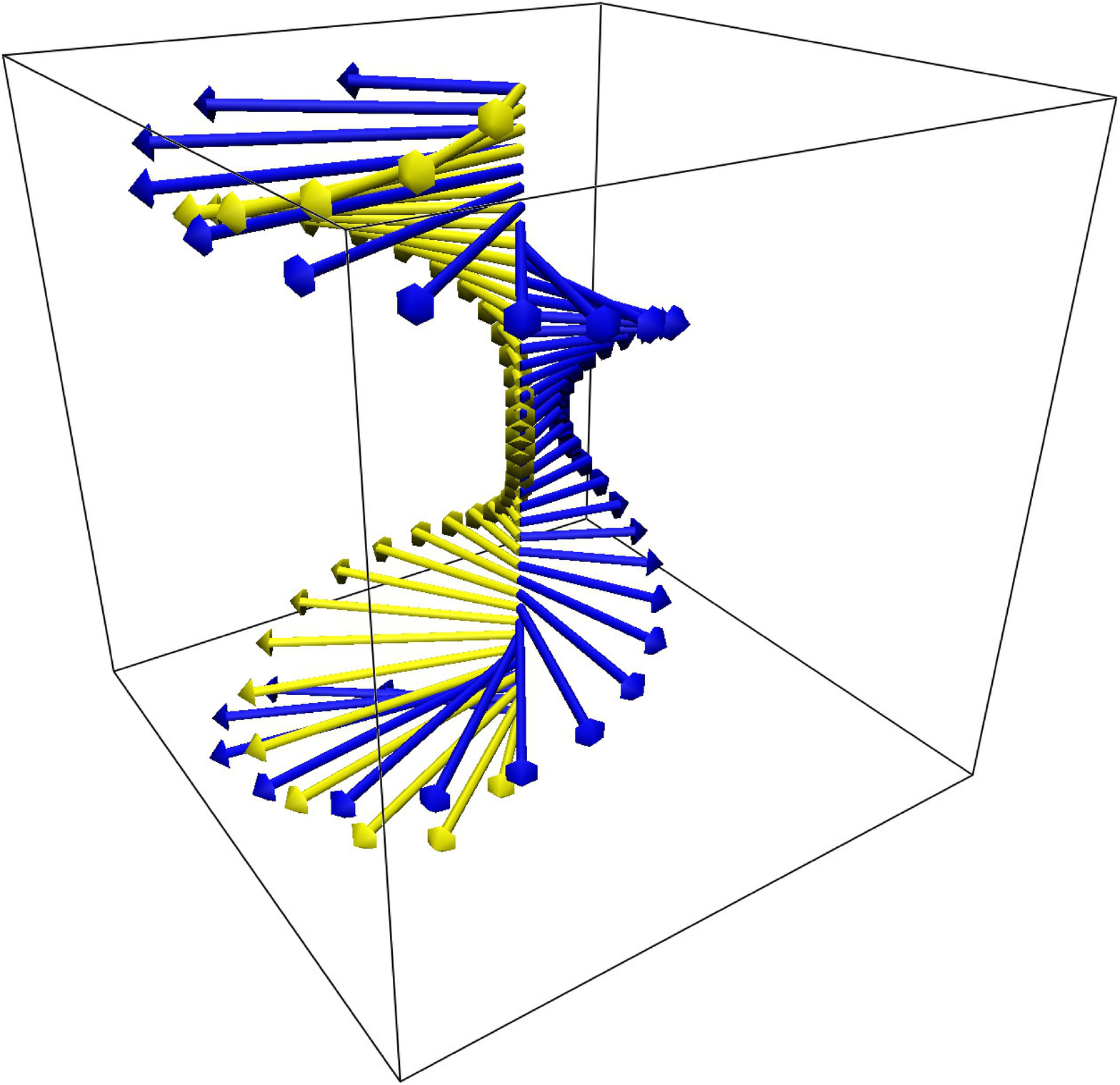}}                       
  \end{center}
\caption{Mean electromagnetic vectors: (a) mean magnetic field $\mBb$; and (b) emf $\emf = \alpha \mBb$ (yellow)  and mean current density $\mJb$ (blue) for $Pr=1$ and $\Rat=50$.}
\label{F:Bvec}
\end{figure}

Three-dimensional visualizations of the mean electrodynamic vectors are shown in Figure \ref{F:Bvec}. The spiral structure of all the vector fields is evident. We see that $\emf \cdot \mJb = 0$ at $Z=0.5$ where both vectors approach zero and point in nearly antiparallel directions as one approaches this point. In accordance with the mean energy equation \eqref{E:magen}, $\emf$ and $\mJb$ are predominantly aligned in regions where $\mJb^2$ (shown in Figure \ref{F:Bprof}(b)) is largest.

\begin{figure}
  \begin{center}
   \subfloat[]{
      \includegraphics[height=4cm]{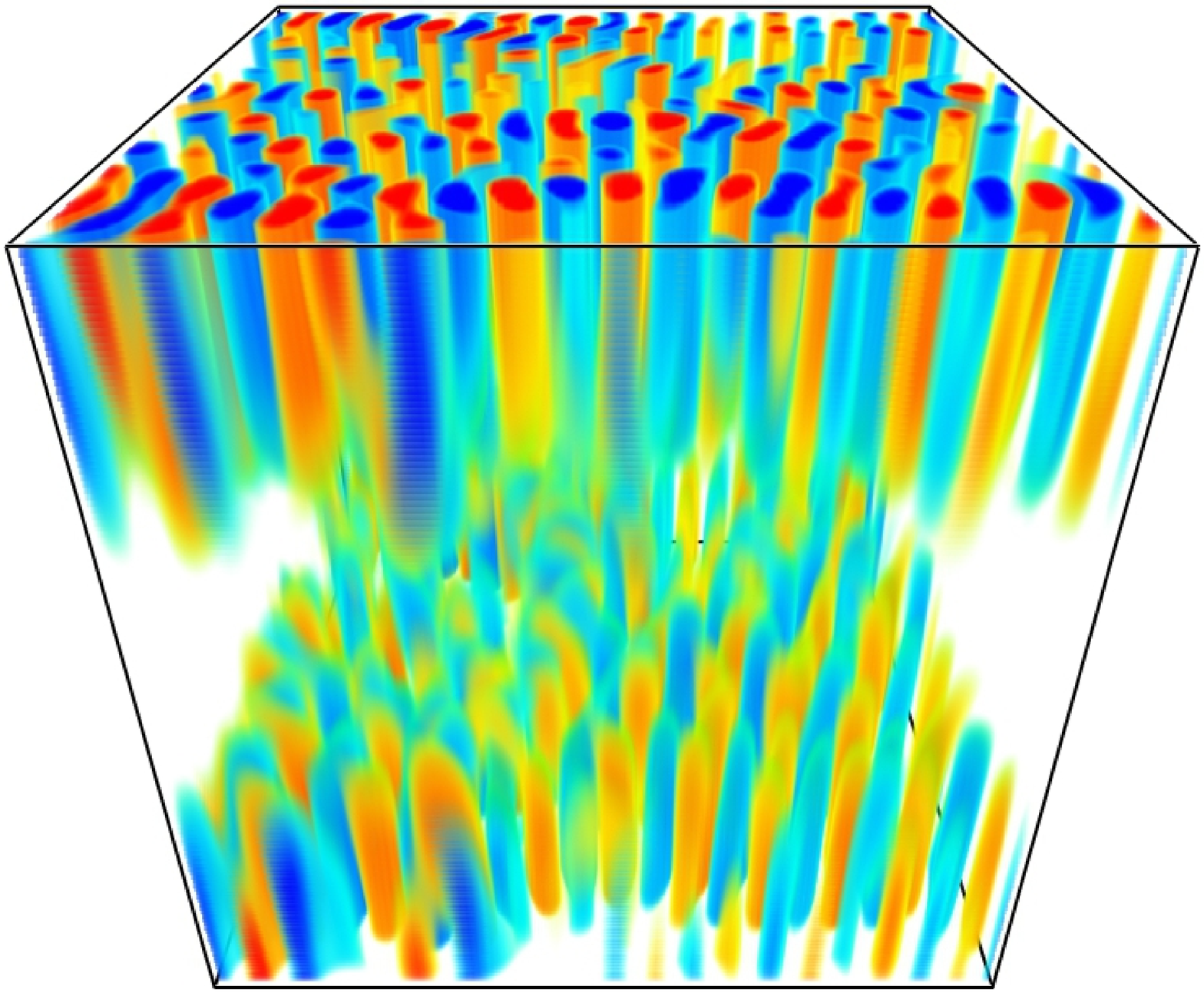}}
      \quad
    \subfloat[]{
      \includegraphics[height=4cm]{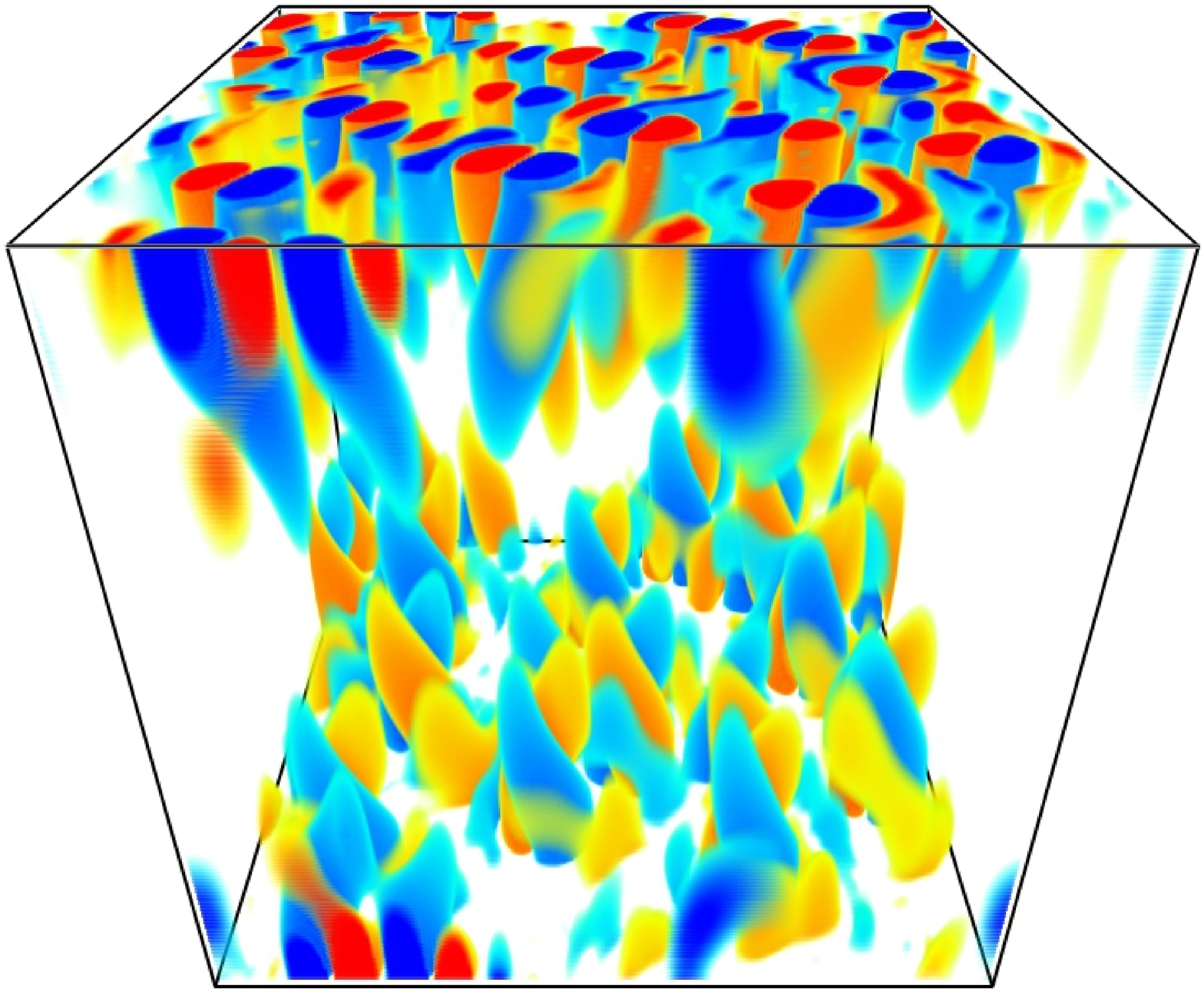}} \\
   \subfloat[]{
      \includegraphics[height=4cm]{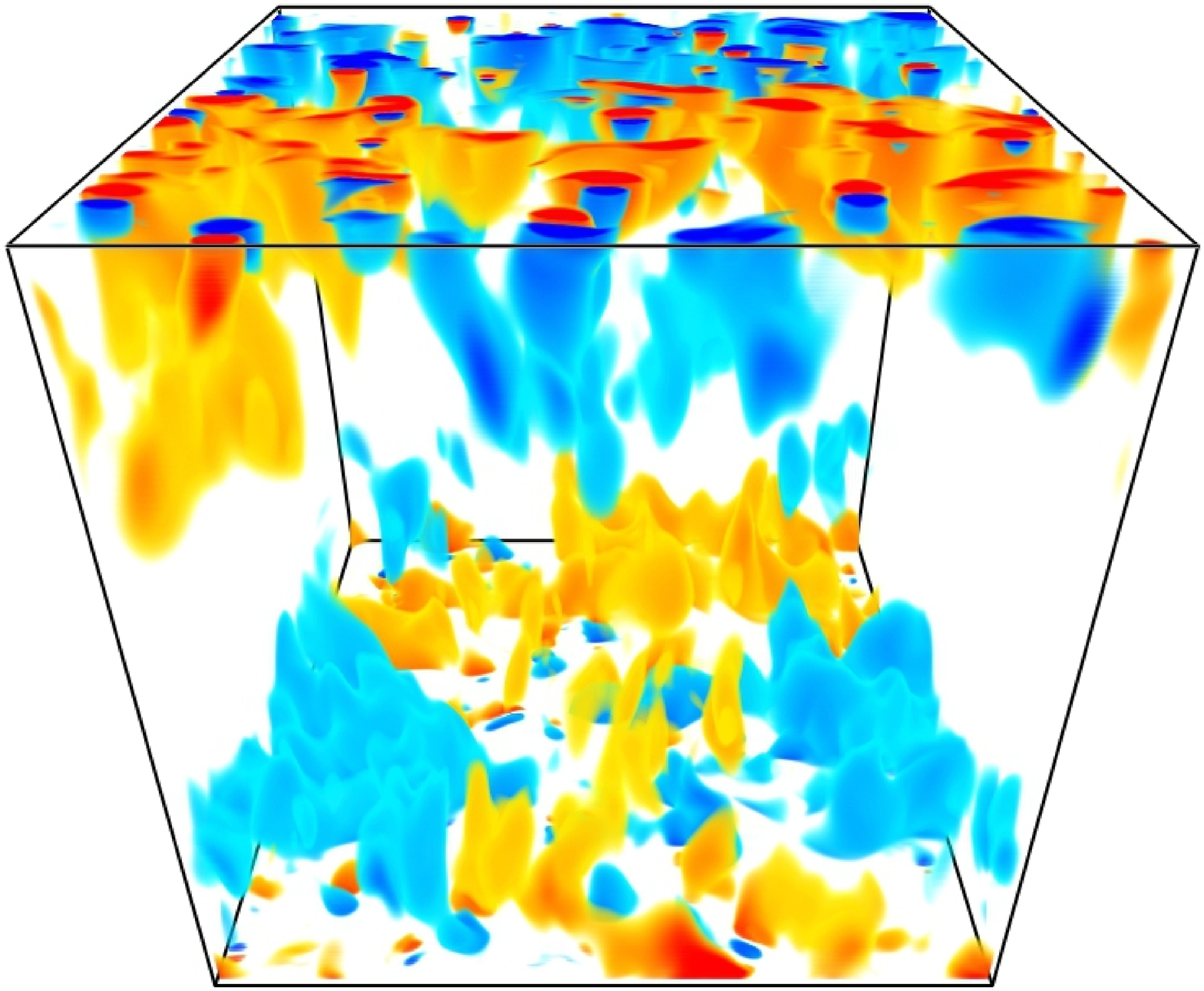}}
      \quad
    \subfloat[]{
      \includegraphics[height=4cm]{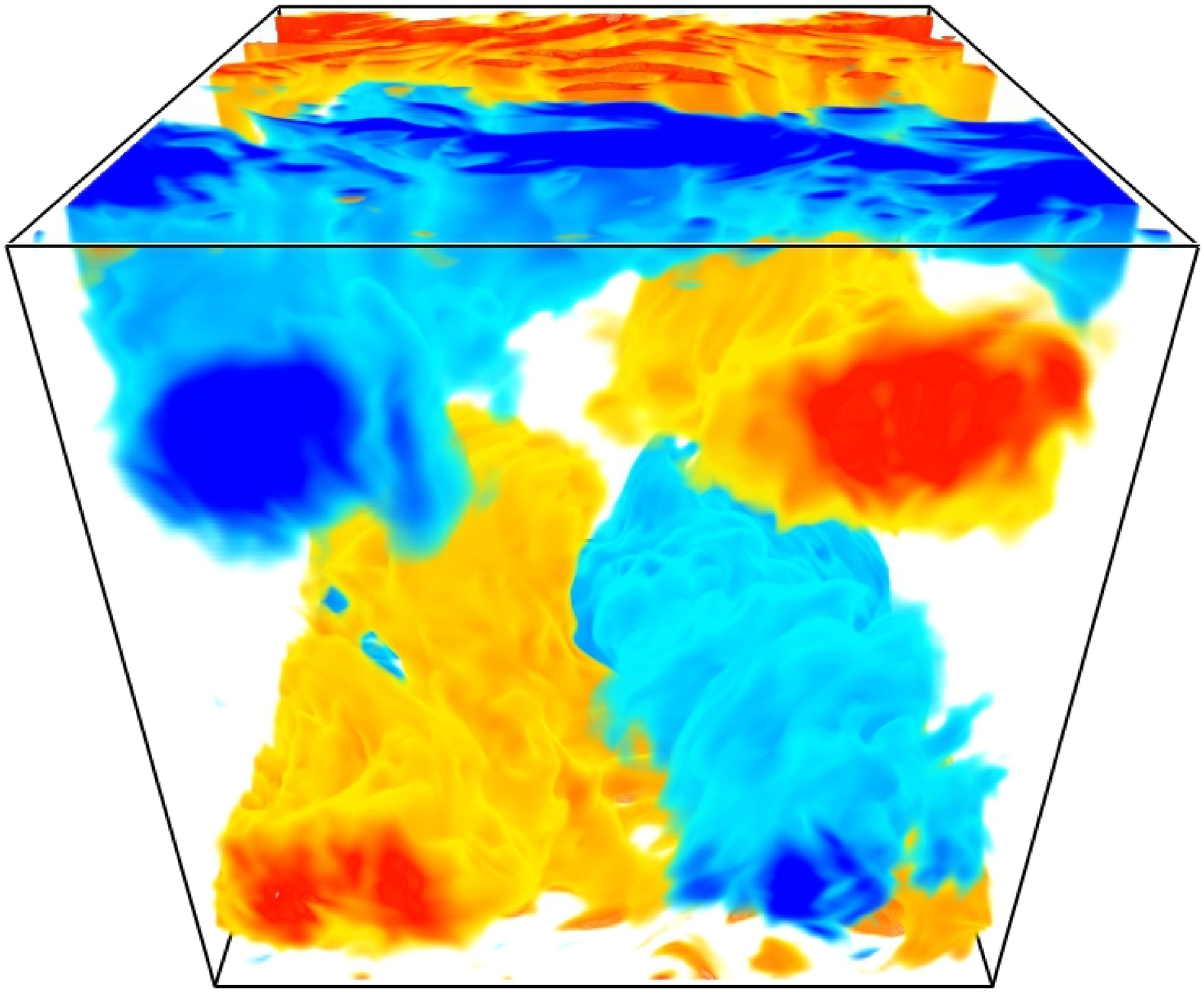}}                              
  \end{center}
\caption{Volumetric renderings of the vertical component of the small-scale current density $\pjz$ for each of the four flow regimes. (a) Cellular regime: $Pr=1$, $\Rat=10$; (b) convective Taylor column regime: $Pr=10$, $\Rat=60$; (c) plume regime: $Pr=10$, $\Rat=200$; (d) geostrophic turbulence regime: $Pr=1$, $\Rat=100$.}
\label{F:jzvol}
\end{figure}

Volumetric renderings of the small-scale vertical current density $\pjz$ are shown in Figure \ref{F:jzvol} for each of the four basic convective flow regimes shown in Figure \ref{F:vortvol}. Taking the vertical component of the curl of equation \eqref{E:finduc1}, and inverting for $\pjz$ yields
\be
\pjz = - \Pmt \, \nabla_\perp^{-2} \, \mBb \cdot \nabla_\perp \zeta' , \label{E:jzinduc}
\ee
showing that $\pjz$ is directly controlled by the structure of $\zeta'$. Of particular note is the trend of increasing scale of $\pjz$ shown in Figure \ref{F:jzvol} as the flow becomes more turbulent with increasing $\Rat$. The presence of the inverse Laplacian in equation \eqref{E:jzinduc} implies that $\pjz$ is sensitive to the presence of large-scale structures in this low $Pm$ limit.

\section{Conclusions}


The characteristics of $\alpha$ and the mean magnetic field are found to be weakly sensitive to the transitions in flow morphology and convective dynamics that occur as the Rayleigh and Prandtl numbers are varied. Indeed, we find qualitatively similar behavior in the dynamo properties for both laminar cellular convection near $\Rat_c$ where $Re \ll 1$ and for $\Rat \gg \Rat_c$ in which $Re \gg 1$. In this latter case an inverse cascade and associated domain-scale vortex is present, which has little influence on the large-scale magnetic field, but does control directly the structure of the small-scale electromagnetic fields. A similar observation was made in the single mode investigation of \citet{mC16}, where, for the fixed-wavenumber solutions, the $\alpha$ profiles became self similar functions of $\Rat$. Though it is merely speculative given the neglect of the Lorentz force in the present work, the relative insensitivity of the large-scale field to the small-scale convective dynamics may suggest why DNS studies, though vastly different in parameter space to natural dynamos, can closely approximate many of the observed properties of the geodynamo and other planetary dynamos \citep[e.g.][]{cJ11b}. 

We find a decrease in the efficiency of the convection for sustaining magnetic field growth as the Rayleigh number is increased, as characterized by an increasing critical magnetic Reynolds number. A similar trend was observed in the maximum Nusselt number single mode solutions of \citet{mC16}, where the enhancement of large-scale magnetic diffusion due to boundary layers was offered as a possible cause. In the present work this reasoning may also apply, though the boundary layers associated with $\alpha$ are much less pronounced. The rise in critical magnetic Reynolds number might also result from the more disordered states of convection that occur at higher Rayleigh numbers, with a possible quantitative measure of this behavior being the substantial decrease in relative kinetic helicity amplitude with increasing Rayleigh number.

The Prandtl number controls the relative importance of inertia for a given value of the Rayleigh number, with higher values of $Pr$ leading to an increased range of $\Rat$ over which inertia plays a weak role in the convective dynamics. For instance, an order of magnitude difference in the Prandtl number leads to an order of magnitude approximate difference in the observed Reynolds numbers at comparable values of the Rayleigh number (e.g.~see Figure \ref{F:Rerms}). This effect of $Pr$ has a direct influence on the critical magnetic Prandtl number required for dynamo action, with lower $Pr$ fluids capable of sustaining dynamo action at lower values of $Pm$ (or $\Pmt$). These findings suggest that compositional convection, which is characterized by $Pr \sim O(100)$ \citep[e.g.][]{mP13}, requires very large values of $\Rat$ to drive a dynamo in comparison to liquid metal thermal convection which is characterized by $Pr \sim 10^{-2}$. 



The QGDM is an \textit{asymptotic} mean field model, and offers the possibility of simulating magnetohydrodynamic flows that are not currently accessible with DNS. No ad-hoc assumptions are necessary to parameterize the $\alpha$-effect since the equations are mathematically closed through the asymptotic reduction procedure. Moreover, these models allow for a more direct appraisal of dynamo physics given the simplified set of equations. Of course, it is necessary to extend the present model to the more realistic case of spherical geometry to allow direct comparison with geo- and planetary magnetic field observations.  The three-dimensional cylindrical annulus model developed by \citet{mC13} offers an intermediate stage between the development of a global spherical model and the plane layer model investigated in the present work.

The asymptotic kinematic dynamo model for rapidly rotating low $Pm$ convection provides much insight into the structure and morphology of magnetic fields at the onset of dynamo action. The observed relative insensitivity of the large-scale magnetic field to changes in flow regime may provide a strategy for parameterizing the effects of small-scale, low Rossby number convection for the purpose of simulating only the large-scale dynamo behavior. However, it is clearly necessary to extend these results into the nonlinear regime in which the Lorentz force is included and the magnetic field can feed back onto the convective dynamics, and to investigate the influence of large-scale horizontal modulations that will allow for a fully multiscale representation of the mean magnetic field \citep[e.g.~see][]{mC15b}. Of particular importance is to determine the behavior of these nonlinear, multiscale solutions and their associated influence on the properties of $\alpha$, and whether subcritical dynamo action can be found \citep[e.g.][]{cJ00,sS04}.




\section*{Acknowledgments}
This work was supported by the National Science Foundation under award numbers EAR-1320991 (MAC and KJ), EAR CSEDI-1067944 (KJ) and DMS EXTREEMS 1407340 (LL).  Volumetric rendering was performed with the visualization software VAPOR \citep{jC05,jC07}.  This work utilized the Janus supercomputer, which is supported by the National Science Foundation (award number CNS-0821794) and the University of Colorado Boulder. The Janus supercomputer is a joint effort of the University of Colorado Boulder, the University of Colorado Denver and the National Center for Atmospheric Research.

%
%
%

\bibliography{../Dynamobib}

\end{document}